\documentclass[aps,preprint,superscriptaddress]{revtex4}
\usepackage{amsfonts}
\usepackage{mathrsfs}
\usepackage{graphicx}
\usepackage{amsmath}
\usepackage{amssymb}
\usepackage{bbm}
\usepackage{subfigure}
\usepackage{caption}
\usepackage{float}
\usepackage{epsfig}
\usepackage{graphicx}
\usepackage{tikz}
\usepackage{color}
\usepackage{physics}
\usepackage{indentfirst}
\usepackage{hyperref}
\usepackage{setspace}
\usepackage{braket}
\usepackage{slashed}
\usepackage{multirow}
\usepackage{diagbox}
\usepackage{cancel}
\usepackage{caption}
\usepackage{array}
\usepackage{ulem}
\newcommand{\dslah}{\slash \! \! \! }

\newcommand{\mi}{\mathrm{i}}

\allowdisplaybreaks[4]

\begin{document}

\title{Spectrum of Light Hexaquark States in Triquark-antitriquark Configuration}
\author{Xuan-Heng Zhang}
\email{zhangxuanheng22@mails.ucas.ac.cn, contact author}
\affiliation{School of Physical Sciences, University of Chinese Academy of Sciences,Yuquan Road 19A, Beijing 100049, China}
\author{Sheng-Qi Zhang}
\email{zhangshengqi20@mails.ucas.ac.cn}
\affiliation{School of Physical Sciences, University of Chinese Academy of Sciences,Yuquan Road 19A, Beijing 100049, China}
\affiliation{Center for High Energy Physics, Peking University, Beijing 100871, China}
\author{Cong-Feng Qiao}
\email{qiaocf@ucas.ac.cn, corresponding author}
\affiliation{School of Physical Sciences, University of Chinese Academy of Sciences,Yuquan Road 19A, Beijing 100049, China}
\affiliation{International Centre for Theoretical Physics Asia-Pacific, University of Chinese Academy of Sciences, Beijing 100190, China}

%\affiliation{~\\}

\begin{abstract}
To understand the nature of $X(2075)$ and $X(2085)$ observed by the BESIII Collaboration in the $p\bar{\Lambda}$ system, we systematically investigate the possibility that these states are compact hexaquark with triquark-antitriquark configurations for the first time. Within the framework of QCD sum rules, the mass spectrum and decay constants of such hexaquark states with quantum numbers $J^P=0^-, 0^+, 1^-, 1^+$ are studied. Consequently, six independent and nondegenerate hexaquark candidates are obtained, among which two $J^P = 1^-$ states exhibit masses consistent with $X(2075)$, while the two $J^P = 1^+$ states differ markedly from the mass of $X(2075)$ or $X(2085)$. The remaining two states with $J^P = 0^+$ and $0^-$ may serve as predictions for potential compact hexaquark configurations. Furthermore, the possible decay modes of these hexaquark states are analyzed, which could be the experimental signatures for their identification. 
\end{abstract}

%\keywords{}

\maketitle

\newpage

\section{Introduction}
The discovery of the $X(3872)$ resonance at the $D^0\bar{D}^{*0}$ threshold by the Belle II Collaboration in 2003~\cite{Belle:2003nnu} inaugurated a new era in hadron spectroscopy, prompting extensive experimental and theoretical efforts that have uncovered a growing number of candidate exotic states. Systematic identification and characterization of these states are essential because they furnish stringent tests of quantum chromodynamics (QCD) in the nonperturbative regime and illuminate the mechanisms of color confinement and multiquark dynamics. Therefore, continued discovery and detailed study of exotic hadronic states remain central goals for both experiment and theory.

Over the past two decades, numerous tetraquark and pentaquark candidates have been firmly established~\cite{LHCb:2015yax,Chen:2015moa,Wang:2015epa,BESIII:2013ris}. Against this backdrop, conducting the exploration for hexaquark system is a natural and necessary extension of hadron spectroscopy. The investigation of hexaquark states has a long-standing history. As early as the 1940s, proposals were advanced to interpret the pion as composite systems formed from nucleon-antinucleon pairs, reflecting early interest in multiquark configurations. More recently, the observations of a series of states near the $p\bar{p}$ threshold, such as $X(1860)$~\cite{BES:2003aic}, $X(1835)$~\cite{BES:2005ega}, $X(1840)$~\cite{BESIII:2013sbm}, and $X(1880)$~\cite{BESIII:2023vvr}, have rekindled and intensified efforts to search for and characterize hexaquark candidates~\cite{Bashkanov:2013cla,Shanahan:2011su,Qiao:2005av,Qiao:2007ce,Wan:2021vny,Wang:2006sna,Wan:2019ake,Chen:2016ymy,Wang:2021qmn,Zhang:2024ulk,Ahmadi:2025vjv,Wan:2025zau,Ikeda:2007nz,Di:2019jsx,Wang:2022jvk,Wan:2025xhf}. Current theoretical investigations for hexaquark states have mainly focused on its molecular configuration, while studies for compact hexaquark states remain comparatively scarce. However, to gain a complete understanding of exotic hadron states, it is essential to explore compact multiquark configurations that complement molecular states and may even be more fundamental.

The arrangements of a hexaquark system can include dibaryon (bound states of two baryons), baryonium (bound states of baryon–antibaryon), and bound states formed by three mesons. Among the baryoniums, a special topology is that the baryon and the antibaryon are not particle-antiparticle conjugates, such as the proton-antihyperon ($p$-$\bar{\Lambda}$) bound state, which lead to the existence of a broader range of exotic states. Experimental evidence for this novel arrangement includes the resonances $X(2075)$ and $X(2085)$, which are both observed in the $p\bar{\Lambda}$ final state \cite{BES:2004fgd, BESIII:2023kgz}. The mass spectra of the $p\bar{\Lambda}$ and $p\bar{\Sigma}$ molecular states are calculated for the possible interpretations of $X(2075)$ and $X(2085)$ in our previous work~\cite{Zhang:2024ulk}. In realistic experimental environments, however, resonance structures are often mixtures of multiple hadron states with nearly degenerate masses and identical quantum numbers. Therefore, it is necessary to investigate other possible configurations in the same mass region. In order to fill the gap in the study of compact hexaquark states, in this work, a compact hexaquark configuration with a triquark-antitriquark structure is constructed, carrying a color configuration $[3_c]$-$[\bar{3}_c]$ and containing one strange quark, which is first proposed. This work aims to explore whether this configuration could contribute to the structures of $X(2075)$ and $X(2085)$.

From the theoretical perspective, this work employs the method of QCD sum rules (QCDSR) \cite{Shifman:1978bx,Shifman:1978by} to investigate exotic states. This approach was proposed in 1979 by Shifman, Vainshtein, and Zakharov (SVZ) based on the first principles of QCD, and it provides an efficient way to incorporate non-perturbative effects of the QCD vacuum. The QCDSR framework allows one to obtain analytical results for hadronic observables such as masses \cite{Shifman:1978bx,Shifman:1978by,Colangelo:2000dp,Narison:2002woh,Zhang:2022obn,Wan:2020fsk,Wan:2020oxt}, decay constants \cite{Dominguez:1987ea,Colangelo:2000dp,Gelhausen:2013wia,Hu:2017dzi,Narison:2020wql}, and form factors \cite{Ioffe:1982qb,Colangelo:2000dp,Khodjamirian:2000ds,Wang:2007ys,Khodjamirian:2020btr,Zhao:2020mod,Neishabouri:2024gbc,Aliev:2010uy,Zhang:2023nxl,Zhang:2024ick,Zhang:2024asb}. For the hexaquark states under consideration in this work, QCDSR has also been shown to provide reliable predictions \cite{Wan:2021vny,Wang:2006sna,Wang:2021qmn}.

In this work, ground-state compact hexaquark configurations of triquark–antitriquark type with quantum numbers $J^P=0^-,\,0^+,\,1^-,\,1^+$ are constructed, yielding in total 24 possible structures. Using QCDSR, their existence, masses, and decay constants are investigated. Also, the results are compared with the experimental data. Furthermore, their possible decay modes are analyzed to provide guidance for experimental identification. The paper is organized as follows. In Sec.\ref{form}, the interpolating currents that couple to these hexaquark states and outline the theoretical framework of QCDSR are constructed. In Sec.\ref{na}, the numerical methods and results are provided. Section \ref{dm} is devoted to the discussion of the possible decay modes of these hexaquark states. Finally, in Sec.\ref{dc}, our predictions are compared with experimental observations, and our findings are summarized.

\section{\label{form}Formalism}
\subsection{Choices of the Currents}
A triquark form $(qq\bar{q})$ that consists of two quarks and one antiquark is considered in this work. According to the decomposition of direct products in the color algebra $SU(3)$,
\begin{equation}
    	[\bar{3}_c]\otimes[3_c]\otimes[3_c]=[15_c]\oplus[\bar{6}_c]\oplus[3_c]\oplus[3_c],
\end{equation}
one can obtain the triquark configurations that form an overall color triplet. Subsequently, according to the reduction of the color $SU(3)$ algebra,
\begin{equation}
    [3_c]\otimes[\bar{3}_c]=[8_c]\oplus[1_c],
\end{equation}
the triquark structure in the color triplet representation $[3_c]$ can combine with the antitriquark structure in the color antitriplet representation $[\bar{3}_c]$ to form a color-singlet bound hexaquark state.

Firstly, the triquark current should be constructed. Following the above discussion, the interpolating current for the triquark in the color triplet representation in the following form is constructed.
\begin{equation}
    \eta^{a}=(\bar{q}_b\Gamma_1q^a)\Gamma_2q^b.
\end{equation}
Next, the triquark--antitriquark currents containing two strange quarks are constructed. Following our previous studies on the states $p\bar{\Lambda}$ and $p\bar{\Sigma}$ \cite{Zhang:2024ulk}, the currents are defined as
\begin{equation}
    \eta^a_{\bar{u}du} = (\bar{u}_b \Gamma_1 d^a)\Gamma_2 u^b,
\end{equation}
\begin{equation}
    \eta^a_{\bar{u}ds} = (\bar{u}_b \Gamma_1 d^a)\Gamma_2 s^b,
\end{equation}
\begin{equation}
    \eta^a_{\bar{u}sd} = (\bar{u}_b \Gamma_1 s^a)\Gamma_2 d^b,
\end{equation}
where $\Gamma_1=\mathbbm{1}, \Gamma_2=\mi\gamma_5$ correspond to the type-I currents, while $\Gamma_1=\mi\gamma_5, \Gamma_2=\mathbbm{1}$ correspond to the type-II currents. Based on the above currents, two configurations of color-singlet triquark-antitriquark interpolating currents containing one strange quark are provided:
\begin{itemize}
    \item $[3_c]_{qqq}$-$[\bar{3}_c]_{qqs}$:
    \begin{equation}\label{eq:j1}
        j_{1(\mu)} = \bar{\eta}_{a;\bar{u}ds}\Gamma_{(\mu)}\eta^a_{\bar{u}du};
    \end{equation}
    \item $[3_c]_{qqq}$-$[\bar{3}_c]_{qsq}$:
    \begin{equation}\label{eq:j2}
        j_{2(\mu)} = \bar{\eta}_{a;\bar{u}sd}\Gamma_{(\mu)}\eta^a_{\bar{u}du};
    \end{equation}
\end{itemize}
where $\Gamma_{(\mu)}=\mathbbm{1}, \mi\gamma_{5}, \gamma_{\mu}, \gamma_{\mu}\gamma_5$ correspond to the four quantum numbers $J^P=0^+,0^-,1^-,1^+$, respectively. By further considering both type-I and type-II currents, a total of 16 possible hexaquark states are obtained.

\subsection{Two-point Correlation Functions}
After selecting the interpolating currents \eqref{eq:j1} and \eqref{eq:j2}, the two-point correlation functions can be calculated, which are defined as
\begin{equation}
    \Pi(q^2) = \mi \int \dd^4 x \, \mathrm{e}^{\mi q \cdot x} 
    \bra{\Omega}\mathbb{T}\{j(x), j^{\dagger}(0)\}\ket{\Omega},
\end{equation}
\begin{equation}
    \Pi_{\mu\nu}(q^2) = \mi \int \dd^4 x \, \mathrm{e}^{\mi q \cdot x} 
    \bra{\Omega}\mathbb{T}\{j_{\mu}(x), j_{\nu}^{\dagger}(0)\}\ket{\Omega},
\end{equation}
where $j(x)$ and $j_{\mu}(x)$ denote the interpolating currents corresponding to the hexaquark states with $J=0$ and $J=1$, respectively, and $\ket{\Omega}$ represents the physical QCD vacuum.  

The two-point correlation function of the tensor type $\Pi_{\mu\nu}(q^2)$ contains contributions from both spin-$0$ and spin-$1$ degrees of freedom and can be decomposed as
\begin{equation}
    \Pi_{\mu\nu}(q^2) = 
    -\left(g_{\mu\nu} - \frac{q_{\mu}q_{\nu}}{q^2}\right)\Pi_1(q^2) 
    + \frac{q_{\mu}q_{\nu}}{q^2}\Pi_0(q^2),
\end{equation}
where the subscripts $1$ and $0$ correspond to spin-$1$ and spin-$0$ states, respectively.  By applying a projection, the spin-$0$ contribution can be removed, obtaining
\begin{equation}
    \Pi_1(q^2) = -\frac{1}{3}\left(g^{\mu\nu} - \frac{q^{\mu}q^{\nu}}{q^2}\right)\Pi_{\mu\nu}(q^2),
\end{equation}
which corresponds to the two-point correlation function of the $J=1$ hexaquark states.

\subsubsection{OPE Side}
The correlation functions for the three configurations can be explicitly expressed as follows based on the Wick’s theorem:
\begin{itemize}
    \item $[3_c]_{qqq}$-$[\bar{3}_c]_{qqs}$:
    \begin{equation}\label{eq:Pi1}
        \begin{aligned}
            \Pi_{1(\mu\nu)}(q^2) = \int[\mathcal{D}x] \, 
            &\Tr\big[\mathcal{S}^{c_1c}_s(-x)\Gamma_2\Gamma_{(\mu)}\Gamma_2
            \mathcal{S}^{bb_1}_q(x)\Gamma_2\Gamma_{(\nu)}\Gamma_2\big] \\
            &\times \Tr\big[\mathcal{S}^{a_1a}_q(-x)\Gamma_1
            \mathcal{S}^{cc_1}_q(x)\Gamma_1\big]
            \Tr\big[\mathcal{S}^{aa_1}_q(x)\Gamma_1
            \mathcal{S}^{b_1b}_q(-x)\Gamma_1\big];
        \end{aligned}
    \end{equation}

    \item $[3_c]_{qqq}$-$[\bar{3}_c]_{qsq}$:
    \begin{equation}\label{eq:Pi2}
        \begin{aligned}
            \Pi_{2(\mu\nu)}(q^2) = \int[\mathcal{D}x] \, 
            &\Tr\big[\mathcal{S}^{c_1c}_q(-x)\Gamma_2\Gamma_{(\mu)}\Gamma_2
            \mathcal{S}^{bb_1}_q(x)\Gamma_2\Gamma_{(\nu)}\Gamma_2\big] \\
            &\times \Tr\big[\mathcal{S}^{a_1a}_s(-x)\Gamma_1
            \mathcal{S}^{cc_1}_q(x)\Gamma_1\big]
            \Tr\big[\mathcal{S}^{aa_1}_q(x)\Gamma_1
            \mathcal{S}^{b_1b}_q(-x)\Gamma_1\big];
        \end{aligned}
    \end{equation}

\end{itemize}
where the shorthand notation $\int[\mathcal{D}x] = -\mi \int \dd^4 x$ has been introduced. In our calculation, the limit $m_u = m_d \to 0$ is taken such that isospin symmetry is preserved, and no distinction is made between the two light flavors $q=u,d$. The quantities $\mathcal{S}_{q/s}^{ab}$ denote the full propagators of the $u,d$ quarks and the $s$ quark, respectively, whose explicit expressions are given as follows:
\begin{equation}
    \begin{aligned}
        \mathrm{i}\mathcal{S}_{q}^{jk}(x) =& \mathrm{i} \delta^{jk} \frac{\dslah{x}}{2\pi^2 x^4} 
- \delta^{jk} m_q \frac{1}{4\pi^2 x^2} 
- \mathrm{i} t^{jk}_a \frac{G^a_{\alpha\beta}}{32 \pi^2 x^2} 
\left( \sigma^{\alpha\beta} \dslah{x} + \dslah{x} \sigma^{\alpha\beta} \right) 
- \delta^{jk} \frac{\langle \bar{q} q \rangle}{12} 
+ \mathrm{i} \delta^{jk} \frac{\dslah{x}}{48} m_q \langle \bar{q} q \rangle 
\\
&- \delta^{jk} \frac{x^2}{192} \langle g_s \bar{q} \sigma \cdot G q \rangle 
+ \mathrm{i} \delta^{jk} \frac{x^2 \dslah{x}}{1152} m_q \langle g_s \bar{q} \sigma \cdot G q \rangle 
- t^{jk}_a \frac{\sigma_{\alpha\beta}}{192} \langle g_s \bar{q} \sigma \cdot G q \rangle\\ 
&- \mathrm{i} t^{jk}_a \frac{1}{768} 
\left( \sigma_{\alpha\beta} \dslah{x} + \dslah{x} \sigma_{\alpha\beta} \right) 
m_q \langle g_s \bar{q} \sigma \cdot G q \rangle-\mi\frac{\dslah x x^2g_s^2\expval{\bar{q}q}^2}{7776}\delta^{jk}-\frac{x^4\expval{\bar{q}q}\expval{g_s^2G^2}}{27648}\delta^{jk}.
\end{aligned}
\label{full-prop}
\end{equation}
Here, both perturbative and nonperturbative contributions at all orders of vacuum condensates are incorporated. For more details on the full propagator can refer to Refs. \cite{Wang:2013vex,Albuquerque:2012jbz}. 

According to the full propagator \eqref{full-prop} and isospin symmetry, the $J^P=0^+$ states of the configurations $[3_c]_{qqq}$-$[\bar{3}_c]_{qqs}$ and $[3_c]_{qqq}$-$[\bar{3}_c]_{qsq}$ for the type-I currents are in fact the same state. Similarly, the $J^P=0^-$ states of these two configurations for the type-II currents are also identical. Their two-point correlation functions are exactly the same and needlessly to be distinguished.  After these considerations, the number of possible nondegenerate hexaquark states whose mass spectra need to be calculated is reduced to 14. The configurations of these two types currents are presented in Table.\ref{currents-config}.

\begin{table}[h]
\centering
\begin{ruledtabular}
\linespread{1.3}\selectfont
\begin{tabular}{ccc}
$J^P$ & Type-I currents & Type-II currents \\
\hline
$0^+$ & $[3_c]_{qqq}$-$[\bar{3}_c]_{qqs}$ & $[3_c]_{qqq}$-$[\bar{3}_c]_{qqs}$, $[3_c]_{qqq}$-$[\bar{3}_c]_{qsq} $ \\
$0^-$ & $[3_c]_{qqq}$-$[\bar{3}_c]_{qqs}$, $[3_c]_{qqq}$-$[\bar{3}_c]_{qsq}$ & $[3_c]_{qqq}$-$[\bar{3}_c]_{qqs}$ \\
$1^-$ & $[3_c]_{qqq}$-$[\bar{3}_c]_{qqs}$, $[3_c]_{qqq}$-$[\bar{3}_c]_{qsq}$ &
$[3_c]_{qqq}$-$[\bar{3}_c]_{qqs}$, $[3_c]_{qqq}$-$[\bar{3}_c]_{qsq}$ \\
$1^+$ & $[3_c]_{qqq}$-$[\bar{3}_c]_{qqs}$, $[3_c]_{qqq}$-$[\bar{3}_c]_{qsq}$ &
$[3_c]_{qqq}$-$[\bar{3}_c]_{qqs}$, $[3_c]_{qqq}$-$[\bar{3}_c]_{qsq}$ \\
\end{tabular}
\end{ruledtabular}
\caption{Independent hexaquark states after considering isospin symmetry and degeneracies.}
\label{currents-config}
\end{table}

Through the Källén-Lehmann spectral representation,
\begin{equation}
    \rho(s) = \frac{1}{\pi} \Im \Pi(s),
\end{equation}
one can correspond the correlation functions given in Eqs.\eqref{eq:Pi1}--\eqref{eq:Pi2} to the spectral density and derive the spectral density in the form of the operator product expansion (OPE), which separates and factorizes the contribution from short distance (Wilson coefficients) and long distance (vacuum condensates). The spectral density in this work are retained up to dimension-13 operators, which can generally be expressed as
    \begin{equation}
\begin{aligned}
    \rho^{\mathrm{OPE}}(s) =& \rho^{\mathrm{pert}}(s) 
+ \rho^{\langle \bar{q} q \rangle}(s) 
+ \rho^{\langle G^2 \rangle}(s) 
+ \rho^{\langle \bar{q} G q \rangle}(s) 
+ \rho^{\langle \bar{q} q \rangle^2}(s) 
+ \rho^{\langle G^3 \rangle}(s)+\rho^{\langle \bar{q} q \rangle \langle G^2 \rangle}(s)  \\
+& \rho^{\langle \bar{q} q \rangle \langle\bar{q}Gq\rangle}(s)+ \rho^{\langle G^4 \rangle}(s) + \rho^{\langle \bar{q} q \rangle^3}(s) + \rho^{\langle \bar{q} G q \rangle \langle G^2 \rangle}(s) 
+ \rho^{\langle \bar{q} q \rangle^2 \langle G^2 \rangle}(s)
+ \rho^{\langle \bar{q} G q \rangle^2}(s)\\
+& \rho^{\langle \bar{q} q \rangle^2 \langle\bar{q}Gq\rangle}(s)+\rho^{\langle \bar{q} q \rangle \langle G^4 \rangle}(s)
+ \rho^{\langle \bar{q} q \rangle^4}(s)+\rho^{\langle \bar{q} q \rangle\langle\bar{q}Gq\rangle \langle G^2 \rangle}(s)+\rho^{\langle \bar{q} q \rangle \langle\bar{q}Gq\rangle^2}(s)\\
+&\rho^{\langle \bar{q}G q \rangle \langle G^4 \rangle}(s).
\end{aligned}
\end{equation}
Subsequently, through the dispersion relation, the spectral density on the OPE side can be used to express the correlation function $\Pi_{X,J^{P}}^{\text{OPE}}(q^2)$ as 
\begin{equation}\label{OPE}
    \Pi_{X,J^{P}}^{\text{OPE}}(q^2) = \int_{s_{\text{min}}}^{\infty} \dd s \ \frac{\rho_{X,J^{P}}^{\text{OPE}}(s)}{s - q^2},
\end{equation}
where $X$ denotes the corresponding ground hadronic state and $J^{P}$ denotes its quantum number; $s_{\text{min}}$ represents the kinematic threshold, typically corresponding to the sum of the masses of all quarks involved in the hadronic interpolating current \cite{Wan:2020oxt,Wan:2020fsk,Wan:2021vny}, i.e. $s_{\text{min}}=m_s^2$ for these hexaquark states with one strange quark. The analytical results of $\rho_{X,J^{P}}^{\text{OPE}}(s)$ are shown in the appendices.

\subsubsection{Phenomenological Side}
In the phenomenological framework, the contributions from the ground state and the excited states (including the continuum spectrum) can be separated as  
\begin{equation}
	\rho_{X,J^{P}}^{\text{Phen}}(s) = \lambda_{X,J^{P}}^2 \, \delta\!\left(s - m_{X,J^{P}}^2\right) 
	+ \theta\!\left(s - s_0\right) \rho_{X,J^{P}}(s),
\end{equation}
where $\lambda_{X,J^{P}}$ and $m_{X,J^{P}}$ denote the decay constant and the mass of the ground state, respectively, and $s_0$ is the threshold parameter, which characterizes the onset of the excited states and the continuum spectrum.  

By applying the dispersion relation, the phenomenological representation of the correlation function can be written as  
\begin{equation}\label{phen}
	\Pi_{X,J^{P}}^{\text{Phen}}(q^2) 
	= \frac{\lambda_{X,J^{P}}^2}{m_{X,J^{P}}^2 - q^2} 
	+ \int_{s_0}^{\infty} \dd s \, \frac{\rho_{X,J^{P}}(s)}{s - q^2},
\end{equation}
where the first term corresponds to the pole contribution of the ground state, while the second term accounts for the contributions from the excited states and the continuum spectrum.

\subsection{Hadronic Mass and Decay Constant}
According to the hypothesis of quark-hadron duality, the correlation functions obtained from the OPE representation and the phenomenological representation should be consistence. In particular, the spectral densities from the two sides are expected to be approximately equal above the continuum threshold parameter $s_0$. Based on this assumption, we can combine Eqs.~\eqref{OPE} and \eqref{phen}. By performing a Borel transformation on both sides of the equation, the contributions from the excited states and the continuum spectrum are exponentially suppressed, leading to
\begin{equation}\label{SR}
	\lambda_{X,J^{P}}^2 \, \mathrm{e}^{-m_{X,J^{P}}^2 / M_B^2} 
	= \int_{s_{\min}}^{s_0} \dd s \, \rho_{X,J^{P}}^{\text{OPE}}(s) \, \mathrm{e}^{-s / M_B^2}.
\end{equation}

From the sum rule given in Eq.~\eqref{SR}, the mass of the ground-state hadron $X$ can be expressed as
\begin{equation}\label{mass}
	m_{X,J^{P}}(s_0, M_B^2) 
	= \sqrt{-\frac{L_{X,J^{P},1}(s_0, M_B^2)}{L_{X,J^{P},0}(s_0, M_B^2)}},
\end{equation}
where
\begin{equation}\label{moment}
	\begin{aligned}
		L_{X,J^{P},0}(s_0, M_B^2) 
		&= \int_{s_{\min}}^{s_0} \dd s \, \rho^{\text{OPE}}(s) \, \mathrm{e}^{-s / M_B^2} 
		+ \Pi^{\text{sum}}(M_B^2), \\
		L_{X,J^{P},1}(s_0, M_B^2) 
		&= \frac{\partial}{\partial (M_B^{-2})} L_{X,J^{P},0}(s_0, M_B^2).
	\end{aligned}
\end{equation}

Here, $\Pi^{\text{sum}}(M_B^2)$ represents the part of the correlation function that originally has no imaginary part but provides nontrivial contribution after the Borel transformation. This contribution is proportional to the masses of the light $u$ and $d$ quarks. Since the light-quark limit $m_u = m_d \to 0$ is taken, this term vanishes in our expressions.  

Furthermore, the decay constant can be extracted from Eq.~\eqref{SR} as
\begin{equation}
	\lambda_{X,J^{P}}(s_0, M_B^2) 
	= \sqrt{\mathrm{e}^{m_{X,J^{P}}^2(s_0, M_B^2)/M_B^2} \, L_{X,J^{P},0}(s_0, M_B^2)}.
\end{equation}

\section{Numerical Analysis\label{na}}
In the numerical calculations of QCDSR, the following input parameters are adopted \cite{Colangelo:2000dp,Ahmadi:2025vjv,Tang:2019nwv,Wan:2021vny}, where $q$ represents the $u,d$ quarks:
\begin{equation}
\begin{array}{ll}
    \langle \bar{q}q \rangle = -(0.24 \pm 0.01)^3 \ \text{GeV}^3, & 
    \langle \bar{s}s \rangle = (1.15 \pm 0.12)\langle \bar{q}q \rangle, \\
    \langle g_s^2 G^2 \rangle = (0.88 \pm 0.25) \ \text{GeV}^4, &
   g_s^2=4\pi\alpha_s=4\pi\times( 0.118 \pm 0.005), \\
    \langle \bar{q}g_s \sigma \cdot G q \rangle = m_0^2 \langle \bar{q}q \rangle, &
    \langle \bar{s}g_s \sigma \cdot G s \rangle = m_0^2 \langle \bar{s}s \rangle, \\
     m_0^2 = (0.8 \pm 0.1) \ \text{GeV}^2, &
    m_s = (95 \pm 5) \ \text{MeV}.
\end{array}
\end{equation}
It should be noted that for $\langle \bar{s}s \rangle$, we have adopted the updated lattice result obtained from Ref. \cite{Davies:2018hmw} . Moreover, in our analytical calculation of the spectral density, we find that the tri-gluon condensate $ \langle g_s^3 G^3 \rangle $ does not contribute to the spectral density, which is consistent with other results for light hexaquark states \cite{Wang:2006sna,Wan:2021vny,Zhang:2024ulk,Ahmadi:2025vjv}. Therefore, we will not provide a numerical value for this condensate.

In constructing the QCDSR, two extra parameters, $s_0$ and $M_B$, are introduced. Since the hadron mass $m_X$ and the decay constant $\lambda_X$ are objective physical observables, their values should not depend on these parameters. Therefore, it is necessary to determine an appropriate parameter region in which the dependence of $m_X$ and $\lambda_X$ on the Borel parameters is minimized, such that the corresponding curves remain as stable as possible. This region is referred to as the Borel window. To identify a suitable Borel window, two assumptions are imposed. First, since the hadronic state under investigation is the ground state, its contribution to the spectral density should be dominant, while the contributions from excited states and the continuum spectra in the high-$s$ region should be as small as possible. According to Eq.\eqref{moment}, the contribution to the spectral density from \( s \gg M_B^2 \) should be significantly suppressed, which was also discussed in Refs.\cite{Chen:2014vha,Azizi:2019xla,Wang:2017sto}. Therefore, the fraction of pole contribution can be defined as 
\begin{equation}
R^\text{PC}_{X,J^{P}} = \frac{L_{X,J^{P},0}(s_0,M_B^2)}{L_{X,J^{P},0}(\infty,M_B^2)}.
\end{equation}
For the hexaquark states, since the spectral density $\rho(s)$ contains higher powers of $s$, the pole contribution is relatively smaller compared to the tetraquark and pentaquark states, which is assumed to be no less than \( 15\%\) \cite{Chen:2014vha,Azizi:2019xla,Wang:2017sto,Wan:2019ake,Wan:2021vny,Wan:2025zau}.  Secondly, the OPE should be convergent, i.e., the contribution of the highest-dimensional condensates to the spectral density should be as small as possible. In this work, we have taken into account the condensate contributions up to dimension 13 operators. The ratio of the dimension 13 condensate contribution is defined as  
\begin{equation}
R^{\text{OPE}}_{X,J^{P}} =\left\vert \frac{L^{\langle O_{13}\rangle}_{X,J^{P},0}(s_0,M_B^2)}{L_{X,J^{P},0}(s_0,M_B^2)}\right\vert,
\end{equation}
where
\begin{equation}
    L^{\langle \mathcal{O}_{13}\rangle}_{X,J^{P},0}(s_0,M_B^2)=\int_{s_{\min}}^{s_0} \dd s \ \rho^{\langle \mathcal{O}_{13}\rangle}(s) \mathrm{e}^{-s / M_B^2} .
\end{equation}

To determine the appropriate range of the threshold parameter $s_0$, the method adopted in Refs.\cite{Wan:2020oxt,Wan:2020fsk,Wan:2021vny,Qiao:2013dda,Tang:2016pcf} is followed. Since $\sqrt{s_0}$ represents the energy scale at which the continuum contribution begins, it should be slightly larger than the ground-state hadron mass $m_X$, i.e.,\(\sqrt{s_0} \sim m_X + \delta \, ,\)
where the typical value of $\delta$ lies within $0.4$--$0.8~\text{GeV}$ \cite{Colangelo:2000dp,Wan:2021vny,Finazzo:2011he}. Within this range of $s_0$, a suitable range of $M_B$ in which the curve of $m_X/\lambda_X - M_B^2$ remains as stable and linear as possible can be searched. The overlapping region of $s_0$ and $M_B$ obtained in this way defines the Borel window, which corresponds to the physically allowed parameter space for the hadronic state under investigation. In practice, an uncertainty of about $\pm 0.1~\text{GeV}$ for $\sqrt{s_0}$ is allowed, while the range of $M_B^2$ should exceed $0.5~\text{GeV}^2$ to ensure sufficient stability of the Borel window.

\begin{figure}[H]
		\centering
		\subfigure[]{\includegraphics[width=0.4\textwidth]{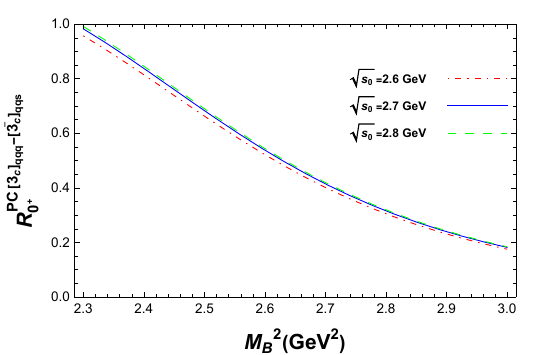}}
		\subfigure[]{\includegraphics[width=0.4\textwidth]{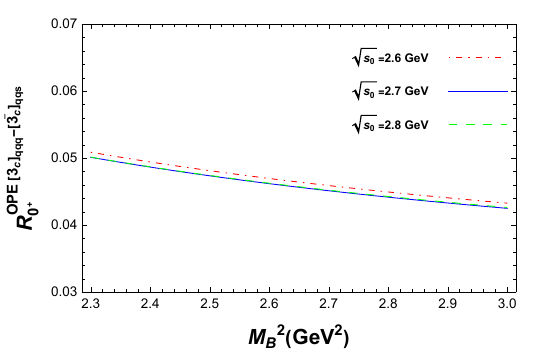}}
		\subfigure[]{\includegraphics[width=0.4\textwidth]{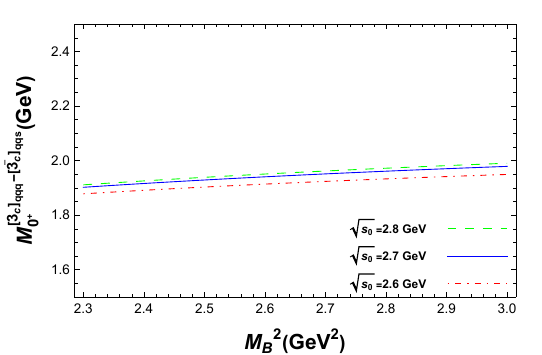}}
		\subfigure[]{\includegraphics[width=0.4\textwidth]{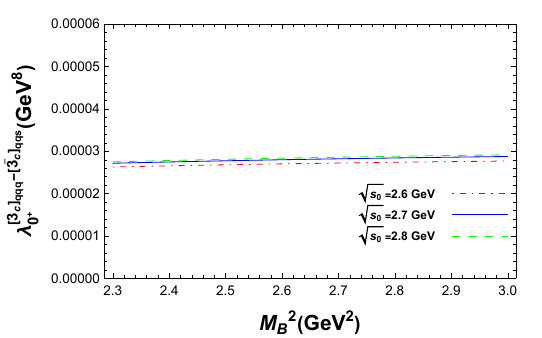}}
		\caption{The figures for $J^P=0^{+}\ [3_c]_{qqq}$-$[\bar{3}_c]_{qqs}$ (or $[3_c]_{qqq}$-$[\bar{3}_c]_{qsq}$) hexaquark states.}
		\label{qqqqqs0+fig}
	\end{figure}

\begin{figure}[H]
		\vspace{-0.5cm}
		\centering
		\subfigure[]{\includegraphics[width=0.4\textwidth]{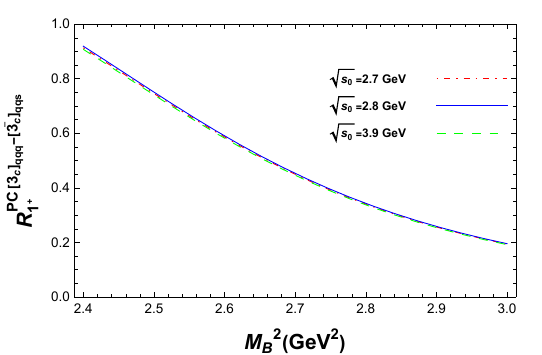}}
		\subfigure[]{\includegraphics[width=0.4\textwidth]{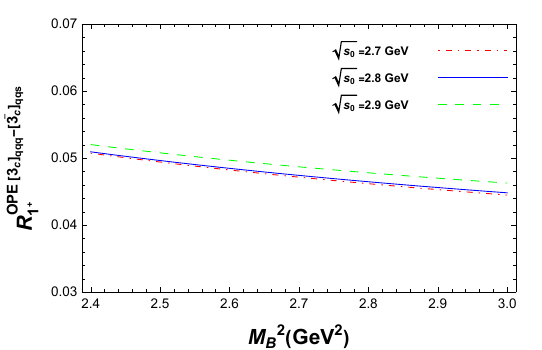}}
		\subfigure[]{\includegraphics[width=0.4\textwidth]{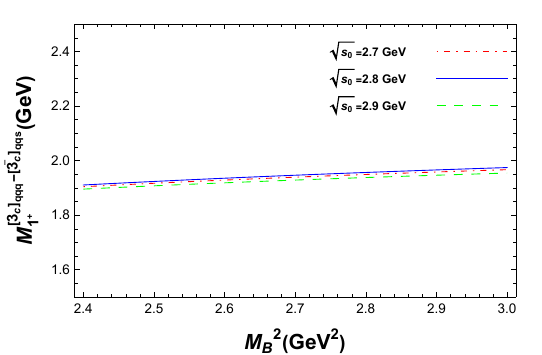}}
		\subfigure[]{\includegraphics[width=0.4\textwidth]{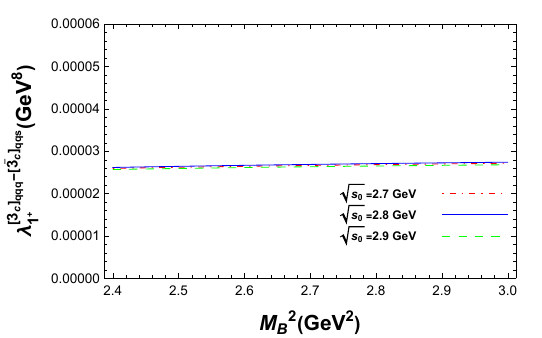}}
		\caption{The figures for $J^P=1^{+}$ $[3_c]_{qqq}$-$[\bar{3}_c]_{qqs}$ hexaquark state. }
		\label{qqqqqs1+fig}
	\end{figure}

\begin{figure}[H]
		\vspace{-0.5cm}
		\centering
		\subfigure[]{\includegraphics[width=0.4\textwidth]{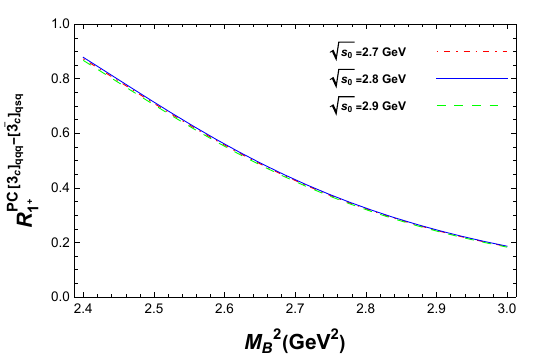}}
		\subfigure[]{\includegraphics[width=0.4\textwidth]{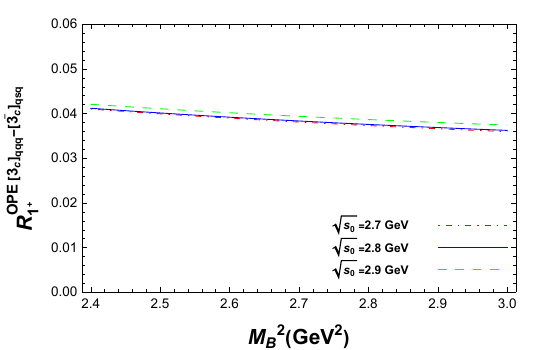}}
		\subfigure[]{\includegraphics[width=0.4\textwidth]{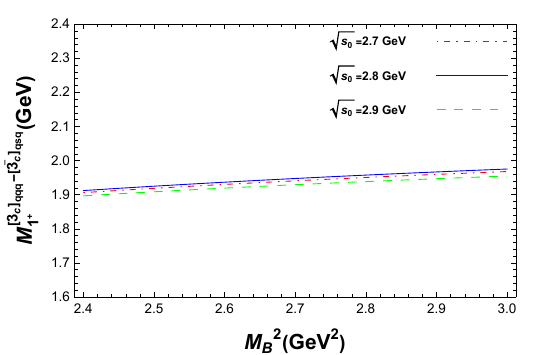}}
		\subfigure[]{\includegraphics[width=0.4\textwidth]{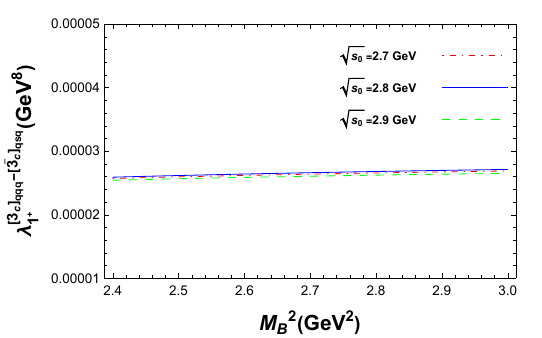}}
		\caption{The figures for $J^P=1^{+}$ $[3_c]_{qqq}$-$[\bar{3}_c]_{qsq}$ hexaquark states. }
		\label{qqqqsq1+fig}
	\end{figure}

\begin{figure}[H]
		\vspace{-0.5cm}
		\centering
		\subfigure[]{\includegraphics[width=0.4\textwidth]{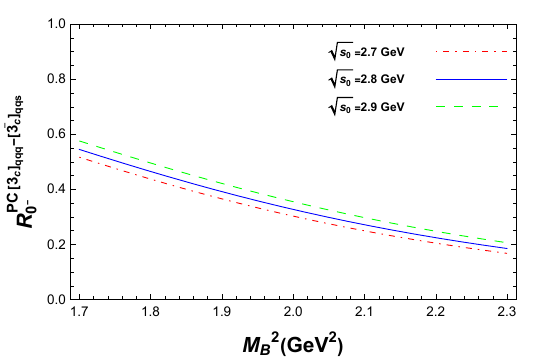}}
		\subfigure[]{\includegraphics[width=0.4\textwidth]{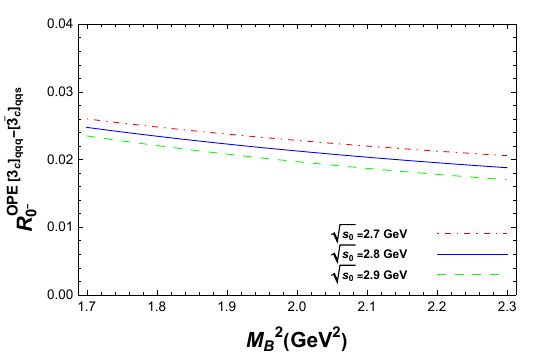}}
		\subfigure[]{\includegraphics[width=0.4\textwidth]{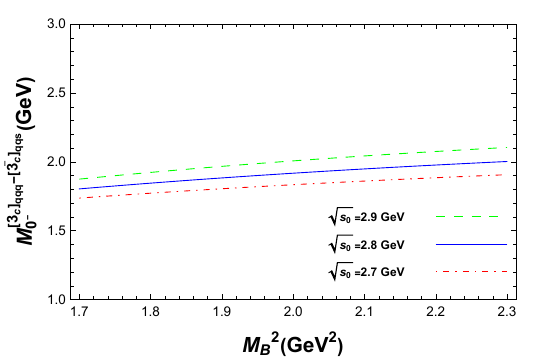}}
		\subfigure[]{\includegraphics[width=0.4\textwidth]{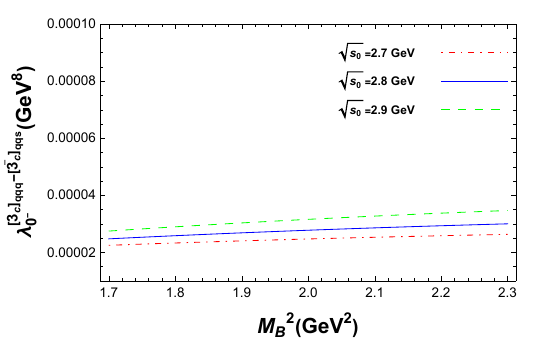}}
		\caption{The figures for $J^P=0^{-}$ $[3_c]_{qqq}$-$[\bar{3}_c]_{qqs}$ (or $[3_c]_{qqq}$-$[\bar{3}_c]_{qsq}$) hexaquark states.}
		\label{qqqqqs0-fig}
	\end{figure}

\begin{figure}[H]
		\vspace{-0.4cm}
		\centering
		\subfigure[]{\includegraphics[width=0.4\textwidth]{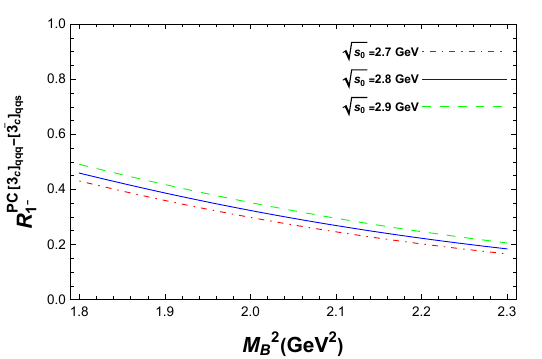}}
		\subfigure[]{\includegraphics[width=0.4\textwidth]{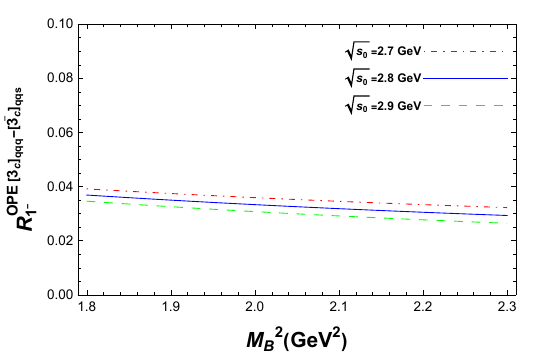}}
		\subfigure[]{\includegraphics[width=0.4\textwidth]{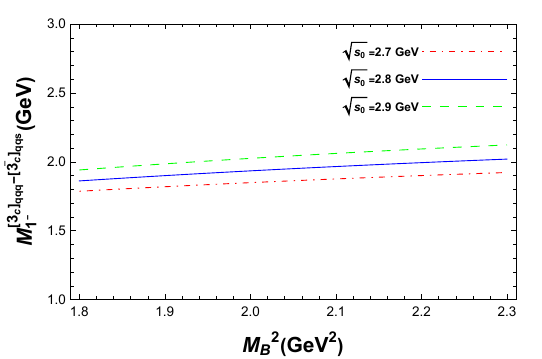}}
		\subfigure[]{\includegraphics[width=0.4\textwidth]{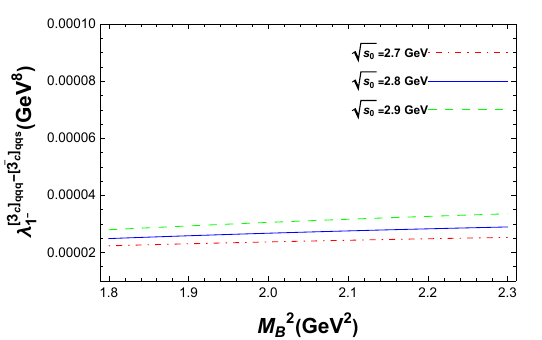}}
		\caption{The figures for $J^P=1^{-}$ $[3_c]_{qqq}$-$[\bar{3}_c]_{qqs}$ hexaquark state. }
		\label{qqqqqs1-fig}
	\end{figure}

\begin{figure}[H]
		\vspace{-0.5cm}
		\centering
		\subfigure[]{\includegraphics[width=0.4\textwidth]{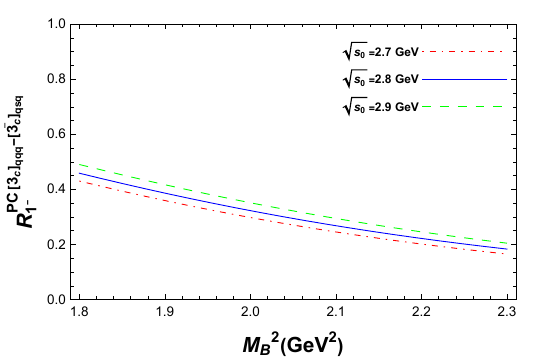}}
		\subfigure[]{\includegraphics[width=0.4\textwidth]{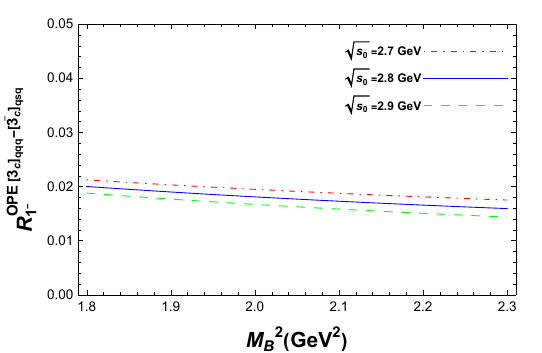}}
		\subfigure[]{\includegraphics[width=0.4\textwidth]{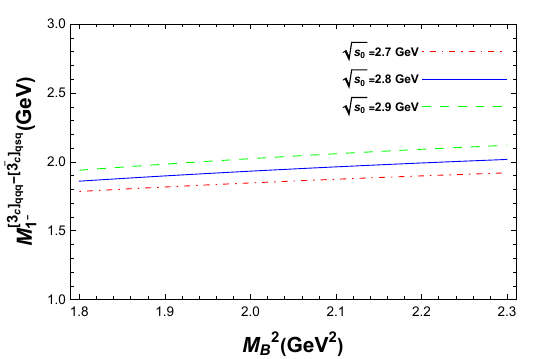}}
		\subfigure[]{\includegraphics[width=0.4\textwidth]{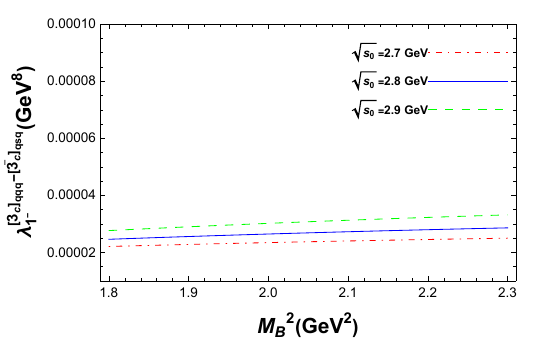}}
		\caption{The figures for $J^P=1^{-}$ $[3_c]_{qqq}$-$[\bar{3}_c]_{qsq}$  hexaquark states. }
		\label{qqqqsq1-fig}
	\end{figure}

As a result, six possible independent hexaquark states can be obtained. These include one $0^+$ state and two $1^+$ states corresponding to the Type-I currents, as well as one $0^-$ state and two $1^-$ states corresponding to the Type-II currents. For the other eight hexaquark states, no stable Borel window can be found regardless of the variation of $s_0$ and $M_B$, indicating that the chosen interpolating currents do not couple well to the corresponding hexaquark configurations. The relevant numerical results are summarized in Table.\ref{tri-antitrire}. Figures \ref{qqqqqs0+fig}--\ref{qqqqsq1-fig} (a)--(d) display the dependence of $R^{\text{PC}}_{X,J^P}$, $R^{\text{OPE}}_{X,J^P}$, the mass $M_{X,J^P}$, and the decay constant $\lambda_{X,J^P}$ on $M_B^2$ for different choices of $s_0$ for these six independent hexaquark states. 

From Table.~\ref{tri-antitrire}, it can be seen that both the masses and the decay constants of the hexaquark states carry certain uncertainties, which of the masses are at the level of $\Lambda_{\text{QCD}}\sim 200 \mathrm{MeV}$. These uncertainties mainly arise from the dependence of the mass on the parameters $s_0$ and $M_B$.
The choice of input parameters, such as the quark masses and the values of the vacuum condensates, also affects these uncertainties. For example, if one adopts the older value from Ref.~\cite{Colangelo:2000dp}, $\langle s\bar{s} \rangle = (0.8 \pm 0.1)\,\langle q\bar{q} \rangle$, the resulting $R^{\text{PC}}_{X,J^P}$ will significantly reduced, leading to a less flat and narrower Borel platform, arising larger uncertainties, which in turn affects the reliability of the results. Besides, for type-II currents, the mass uncertainty is three to four times larger than for type-I currents, but both are of the order $\Lambda_{\text{QCD}} \sim 200 \, \mathrm{MeV}$. This may results in a larger $R^{\text{PC}}_{X,J^P}$ for type-I currents, which suggests that the contribution from excited states and the continuum spectrum is smaller in the type-I spectrum. Hence, the result for the ground state is more reliable, yielding a wider and flatter Borel window with less dependence on $s_0$ and $M_B$, leading to smaller uncertainties.
 
\begin{table}[h]
\linespread{1.3}\selectfont
       \small
        \begin{ruledtabular}
        %\linespread{1.1}\selectfont
		\begin{tabular}{ccccccccc}
			Currents &  $J^{P}$ &  Configurations & $\sqrt{s_0}(\mathrm{GeV})$ & $M_B^2 (\mathrm{GeV}^2)$ & $m_X(\mathrm{GeV})$ & $\lambda_{X}(10^{-5}\mathrm{GeV}^8)$ & $R^{\mathrm{PC}}(\%)$ & $R^{\mathrm{OPE}}(\%)$\\
			\hline
			Type-I  
			& $0^{+}$ & $[3_c]_{qqq}$-$[\bar{3}_c]_{qqs}$ & $2.7\pm 0.1$ & $2.3-3.0$ & $1.93\pm 0.06$ & $2.77\pm 0.14$& $16-98$&$4.3-5.2$ \\
			& $1^{+}$ & $[3_c]_{qqq}$-$[\bar{3}_c]_{qqs}$ & $2.8\pm 0.1$ & $2.4-3.0$ & $1.93\pm 0.04$ & $2.65\pm 0.08$& $19-92$&$4.5-5.4$ \\
			&  & $[3_c]_{qqq}$-$[\bar{3}_c]_{qsq}$ & $2.8\pm 0.1$ & $2.4-3.0$ & $1.94\pm 0.04$ & $2.62\pm 0.03$& $18-88$&$3.6-4.2$ \\
			\hline
			Type-II  & $0^{-}$ & $[3_c]_{qqq}$-$[\bar{3}_c]_{qqs}$ & $2.8\pm 0.1$ & $1.7-2.3$ & $1.92\pm 0.18$& $2.86\pm 0.61$& $17-58$&$1.7-2.6$ \\
			& $1^{-}$ & $[3_c]_{qqq}$-$[\bar{3}_c]_{qqs}$ & $2.8\pm 0.1$ & $1.8-2.3$ & $1.95\pm 0.17$  & $2.79\pm 0.56$ & $17-49$&$2.6-3.9$ \\
			& & $[3_c]_{qqq}$-$[\bar{3}_c]_{qsq}$ & $2.8\pm 0.1$ & $1.8-2.3$ & $1.95\pm 0.17$&$2.76\pm 0.55$ & $16-49$&$1.4-2.1$ \\
		\end{tabular}
        \end{ruledtabular}
		\caption{Numerical results of the six possible independent and nondegenerate hexaquark states.}
		\label{tri-antitrire}
	\end{table}
    
\section{Decay Modes Analyses \label{dm}}
In realistic experimental environments, the observed hadronic states are usually not pure states, but rather mixtures of several hadrons with nearly degenerate masses and identical quantum numbers (such as $J^P$ and strange numbers). To distinguish these states, it is essential to analyze their decay patterns. In this section, the possible decay modes of the hexaquark states obtained in our calculations are provided, with the expectation that ongoing experiments such as BESIII and Belle-II may be able to search for them.

The dominant strong decay channels are summarized in Table \ref{Decay}. Since the central values of the calculated hexaquark masses lie below the thresholds of $p\bar{\Lambda}$ and $p\bar{\Sigma}$, decays into these two baryons are forbidden. However, once uncertainties are taken into account, the $J^P=0^-, 1^-$ hexaquark states may allow for such decay channels. Therefore, these two decay patterns are also listed in parentheses in Table \ref{Decay}. Furthermore, the $p\bar{\Lambda}$ and $p\bar{\Sigma}$ molecular state is more prone to directly decay into $p+\bar{\Lambda}$ or $p+\bar{\Sigma}$ because it consists of two loosely bound baryon-antibaryon clusters and thus is less stable compared to the compact hexaquark state. Decays into three-meson final states are expected to dominate \cite{Zhang:2024ulk,Wan:2021vny}, since the thresholds of these final states are smaller than the masses of hexaquark states. In addition, these two types of hexaquark configurations can also undergo weak decays, such as transitions to a nucleon–-antinucleon pair $N\bar{N}$. However, such processes involve the Cabibbo-suppressed transition $s \to u$, making them much more difficult to observe compared to the strong decays.


    \begin{table}[h]
        \begin{ruledtabular}
            \linespread{1.3}\selectfont
		\begin{tabular}{ccccc}
			$J^{P}$ & $0^{-}$ & $0^{+}$ & $1^{-}$ & $1^+$ \\
			\hline
			$[3_c]_{qqq}$-$[\bar{3}_c]_{qqs}$
			& $\pi\pi K$, $\omega\omega K$ & $\pi\pi K^{*}$, $\omega\omega K^{*}$ & $\pi\pi K^{*}$,  $\omega\omega K^{*}$  & $\pi\pi K^{*}$,   $\omega\omega K^{*}$  \\
            & $(p\bar{\Lambda}/p\bar{\Sigma})$ &$\pi\rho K$   & $\pi\rho K$ $(p\bar{\Lambda}/p\bar{\Sigma})$  &  $\pi\rho K$    \\
            \hline
			$[3_c]_{qqq}$-$[\bar{3}_c]_{qsq}$
			& $\pi\pi K$, $\omega\omega K$ & $\pi\pi K^{*}$, $\omega\omega K^{*}$ & $\pi\pi K^{*}$,  $\omega\omega K^{*}$  & $\pi\pi K^{*}$,   $\omega\omega K^{*}$ \\
            &$(p\bar{\Lambda}/p\bar{\Sigma})$  &$\pi\rho K$   & $\pi\rho K$ $(p\bar{\Lambda}/p\bar{\Sigma})$  &  $\pi\rho K$    
		\end{tabular}
        \end{ruledtabular}
	\caption{Typical decay modes of the hexaquark states for each quantum number.}
	\label{Decay}
\end{table}

\section{Discussion and Conclusions\label{dc}}

In this work, we explore the compact hexaquark states constructed from two different triquark currents, in which the only two possible configurations, $[3_c]_{qqq}$-$[\bar{3}_c]_{qqs}$ and $[3_c]_{qqq}$-$[\bar{3}_c]_{qsq}$ with quantum numbers $J^P=0^-,0^+,1^-,1^+$ are taken into account. Their masses and decay constants are calculated within the QCDSR framework. The results indicate the probable existence of six independent hexaquark states. The possible decay modes of these six hexaquark states are analyzed, including both strong and weak decay channels. These predicted decay patterns provide a theoretical basis for their potential identification in ongoing experiments such as BESIII and Belle-II.

Phenomenologically, one can find that the mass of $X(2075)$ lies within the predicted range of the $J^P=1^-$ hexaquark states in $[3_c]_{qqq}$-$[\bar{3}_c]_{qqs}$ and $[3_c]_{qqq}$-$[\bar{3}_c]_{qsq}$ configurations, suggesting that $X(2075)$ may contain such hexaquark components. In contrast, the mass of $X(2085)$ does not fall into the predicted region for the $J^P=1^+$ hexaquark states considered here, implying that its internal structure cannot be explained by the present configurations alone. Our previous work \cite{Zhang:2024ulk} showed that no $J^P = 1^+$ molecular states of $p\bar{\Lambda}$ and $p\bar{\Sigma}$ can be identified within the QCDSR framework. In contrast, in the present work we find compact hexaquark states with the same quantum numbers. This suggests that the $J^P = 1^+$ state $X(2085)$ is more likely to have a compact hexaquark structure rather than a molecular configuration. To clarify the nature of $X(2085)$, a more comprehensive analysis is desirable that includes a mixture of interpolating currents in Eqs.~\eqref{eq:j1}–\eqref{eq:j2} and other possible configurations, such as currents in Ref. \cite{Zhang:2024ulk}. Moreover, the two compact hexaquark states predicted in this work with quantum numbers $J^P = 0^+$ and $J^P = 0^-$ can serve as candidate for the open strange hadrons near 2 GeV.

Finally, this study provides a further theoretical insight to the spectroscopy of multiquark system. The molecular and multiquark structures may coexist, which pose a challenge to us in deciphering the exotic states.

\begin{acknowledgments}
 We appreciate the enlightening discussion with Bing-Dong Wan and Liang Tang. This work was supported in part by the National Key Research and Development Program of China under Contracts No. 2020YFA0406400, by the National Natural Science Foundation of China (NSFC) under the Grants 12475087 , 12235008 and 12547114. 
 
\end{acknowledgments}
\bibliography{references.bib}

\newpage
\begin{appendix}
In the appendix, we will present the analytical results for the spectral densities corresponding to the 14 scenarios in Table.\ref{currents-config}. During the calculation of the spectral densities, the \texttt{FeynCalc} package     \cite{Shtabovenko:2020gxv,Shtabovenko:2016sxi,Mertig:1990an} was utilized to produce the traces of the $\gamma$-matrices. 

We will expand the spectral densities as
\begin{equation}
    \rho^{\text{OPE}}=\rho^{\text{pert}}+\sum_{n=3}^{13}\rho^{\langle\mathcal{O}_n\rangle},
\end{equation}
where $n$ denotes for the dimension of the vacuum condensates.

\section{Spectra of Type-I Currents}

\subsection{$0^{+}$ $[3_c]_{qqq}$-$[\bar{3}_c]_{qqs}$}
\vspace{-1.5cm}
\begin{eqnarray}
         \rho^{\text{pert}}&=&\frac{s^7}{123312537600 \pi ^{10}},\\
          \rho^{\langle\mathcal{O}_3\rangle}&=&\frac{m_s s^5 (2 \expval{q\bar{q}}+\expval{s\bar{s}})}{78643200 \pi ^8},\\
          \rho^{\langle\mathcal{O}_4\rangle}&=&0,\\
           \rho^{\langle\mathcal{O}_5\rangle}&=&-\frac{m_s s^4 (3 \expval{qG\bar{q}}+\expval{sG\bar{s}})}{15728640 \pi ^8},\\
           \rho^{\langle\mathcal{O}_6\rangle}&=&-\frac{\expval{q\bar{q}} s^4 (2 \expval{q\bar{q}}+\expval{s\bar{s}})}{983040 \pi ^6}+\frac{g_s^2 s^4 \left(5 \expval{q\bar{q}}^2+\expval{s\bar{s}}^2\right)}{106168320 \pi ^8},\\
           \rho^{\langle\mathcal{O}_7\rangle}&=&\frac{m_s \expval{q\bar{q}}  \expval{g_s^2G^2}  s^3}{9437184 \pi ^8},\\
           \rho^{\langle\mathcal{O}_8\rangle}&=&\frac{s^3 (\expval{q\bar{q}} (4 \expval{qG\bar{q}}+\expval{sG\bar{s}})+\expval{s\bar{s}} \expval{qG\bar{q}})}{196608 \pi ^6},\\
           \rho^{\langle\mathcal{O}_9\rangle}&=&-\frac{m_s \expval{q\bar{q}}^2 s^2 (2 \expval{q\bar{q}}+\expval{s\bar{s}})}{3072 \pi ^4}+\frac{g_s^2 m_s \expval{q\bar{q}}^2 s^2 (8 \expval{q\bar{q}}+5 \expval{s\bar{s}})}{663552 \pi ^6},\\
            \rho^{\langle\mathcal{O}_{10}\rangle}&=&-\frac{\expval{q\bar{q}}  \expval{g_s^2G^2}  s^2 (2 \expval{q\bar{q}}+\expval{s\bar{s}})}{294912 \pi ^6}-\frac{\expval{qG\bar{q}} s^2 (2 \expval{qG\bar{q}}+\expval{sG\bar{s}})}{65536 \pi ^6},\\
            \rho^{\langle\mathcal{O}_{11}\rangle}&=&\frac{m_s \expval{q\bar{q}} s (\expval{q\bar{q}} (9 \expval{qG\bar{q}}+\expval{sG\bar{s}})+3 \expval{s\bar{s}} \expval{sG\bar{s}})}{3072 \pi ^4}\\
            & &-\frac{g_s^2 m_s \expval{q\bar{q}}^2 s (12 \expval{qG\bar{q}}+5 \expval{sG\bar{s}})}{663552 \pi ^6},\\
           \rho^{\langle\mathcal{O}_{12}\rangle}&=&\frac{\expval{q\bar{q}}^3 s (\expval{q\bar{q}}+2 \expval{s\bar{s}})}{384 \pi ^2}+\frac{5 g_s ^4 \expval{q\bar{q}}^2 s \left(2 \expval{q\bar{q}}^2+\expval{s\bar{s}}^2\right)}{4478976 \pi ^6}\\
           & &-\frac{g_s^2 \expval{q\bar{q}}^2 s \left(3 \expval{q\bar{q}}^2+2 \expval{q\bar{q}} \expval{s\bar{s}}+\expval{s\bar{s}}^2\right)}{20736 \pi ^4},\\
              \rho^{\langle\mathcal{O}_{13}\rangle}&=&-\frac{m_s \expval{qG\bar{q}} (18 \expval{q\bar{q}} \expval{qG\bar{q}}+4 \expval{q\bar{q}} \expval{sG\bar{s}}+3 \expval{s\bar{s}} \expval{qG\bar{q}})}{12288 \pi ^4}.
     \end{eqnarray}

\subsection{$0^{-}$ $[3_c]_{qqq}$-$[\bar{3}_c]_{qqs}$}
\vspace{-1.5cm}
\begin{eqnarray}
         \rho^{\text{pert}}&=&\frac{s^7}{123312537600 \pi ^{10}},\\
          \rho^{\langle\mathcal{O}_3\rangle}&=&-\frac{m_s s^5 (2 \expval{q\bar{q}}-\expval{s\bar{s}})}{78643200 \pi ^8},\\
          \rho^{\langle\mathcal{O}_4\rangle}&=&0,\\
           \rho^{\langle\mathcal{O}_5\rangle}&=&\frac{m_s s^4 (3 \expval{qG\bar{q}}-\expval{sG\bar{s}})}{15728640 \pi ^8},\\
           \rho^{\langle\mathcal{O}_6\rangle}&=&-\frac{\expval{q\bar{q}} s^4 (2 \expval{q\bar{q}}-\expval{s\bar{s}})}{983040 \pi ^6}+\frac{g_s^2 s^4 \left(5 \expval{q\bar{q}}^2+\expval{s\bar{s}}^2\right)}{106168320 \pi ^8},\\
           \rho^{\langle\mathcal{O}_7\rangle}&=&-\frac{m_s \expval{q\bar{q}} \expval{g_s^2G^2} s^3}{9437184 \pi ^8},\\
           \rho^{\langle\mathcal{O}_8\rangle}&=&\frac{s^3 (\expval{q\bar{q}} (4 \expval{qG\bar{q}}-\expval{sG\bar{s}})-\expval{s\bar{s}} \expval{qG\bar{q}})}{196608 \pi ^6},\\
           \rho^{\langle\mathcal{O}_9\rangle}&=&-\frac{m_s \expval{q\bar{q}}^2 s^2 (2 \expval{q\bar{q}}-\expval{s\bar{s}})}{3072 \pi ^4}+\frac{g_s^2 m_s \expval{q\bar{q}}^2 s^2 (5 \expval{s\bar{s}}-8 \expval{q\bar{q}})}{663552 \pi ^6},\\
            \rho^{\langle\mathcal{O}_{10}\rangle}&=&-\frac{\expval{qG\bar{q}} s^2 (2 \expval{qG\bar{q}}-\expval{sG\bar{s}})}{65536 \pi ^6}-\frac{\expval{q\bar{q}} \expval{g_s^2G^2} s^2 (2 \expval{q\bar{q}}-\expval{s\bar{s}})}{294912 \pi ^6},\\
            \rho^{\langle\mathcal{O}_{11}\rangle}&=&-\frac{m_s \expval{q\bar{q}} s (\expval{q\bar{q}} (9 \expval{qG\bar{q}}-\expval{sG\bar{s}})-3 \expval{s\bar{s}} \expval{sG\bar{s}})}{3072 \pi ^4}\\
            & & +\frac{g_s^2 m_s \expval{q\bar{q}}^2 s (12 \expval{qG\bar{q}}-5 \expval{sG\bar{s}})}{663552 \pi ^6},\\
           \rho^{\langle\mathcal{O}_{12}\rangle}&=&\frac{\expval{q\bar{q}}^3 s (\expval{q\bar{q}}-2 \expval{s\bar{s}})}{384 \pi ^2}-\frac{g_s^2 \expval{q\bar{q}}^2 s \left(3 \expval{q\bar{q}}^2-2 \expval{q\bar{q}} \expval{s\bar{s}}+\expval{s\bar{s}}^2\right)}{20736 \pi ^4}\\
           & &+\frac{5 g_s ^4 \expval{q\bar{q}}^2 s \left(2 \expval{q\bar{q}}^2+\expval{s\bar{s}}^2\right)}{4478976 \pi ^6},\\
              \rho^{\langle\mathcal{O}_{13}\rangle}&=&\frac{m_s \expval{qG\bar{q}} (18 \expval{q\bar{q}} \expval{qG\bar{q}}-4 \expval{q\bar{q}} \expval{sG\bar{s}}-3 \expval{s\bar{s}} \expval{qG\bar{q}})}{12288 \pi ^4}.
     \end{eqnarray}

\subsection{$0^{-}$ $[3_c]_{qqq}$-$[\bar{3}_c]_{qsq}$}
\vspace{-1.5cm}
\begin{eqnarray}
         \rho^{\text{pert}}&=&\frac{s^7}{123312537600 \pi ^{10}},\\
          \rho^{\langle\mathcal{O}_3\rangle}&=&\frac{m_s s^5 (2 \expval{q\bar{q}}+\expval{s\bar{s}})}{78643200 \pi ^8},\\
          \rho^{\langle\mathcal{O}_4\rangle}&=& \rho^{\langle\mathcal{O}_{13}\rangle}=0,\\
           \rho^{\langle\mathcal{O}_5\rangle}&=&-\frac{m_s s^4 (3 \expval{qG\bar{q}}+\expval{sG\bar{s}})}{15728640 \pi ^8},\\
           \rho^{\langle\mathcal{O}_6\rangle}&=&-\frac{\expval{q\bar{q}}  \expval{s\bar{s}}s^4}{983040 \pi ^6}+\frac{g_s ^2 s^4 \left(5 \expval{q\bar{q}}^2+\expval{s\bar{s}}^2\right)}{106168320 \pi ^8},\\
           \rho^{\langle\mathcal{O}_7\rangle}&=&\frac{m_s  \expval{q\bar{q}} \expval{g_s^2G^2} s^3}{9437184 \pi ^8},\\
           \rho^{\langle\mathcal{O}_8\rangle}&=&\frac{s^3 \expval{q\bar{q}}\expval{sG\bar{s}}+\expval{s\bar{s}} \expval{qG\bar{q}})}{196608 \pi ^6},\\
           \rho^{\langle\mathcal{O}_9\rangle}&=&\frac{g_s ^2 m_s  \expval{q\bar{q}}^2 s^2 (8 \expval{q\bar{q}}+5 \expval{s\bar{s}})}{663552 \pi ^6},\\
            \rho^{\langle\mathcal{O}_{10}\rangle}&=&-\frac{s^2\expval{qG\bar{q}} \expval{sG\bar{s}}}{65536 \pi ^6}-\frac{\expval{q\bar{q}} \expval{s\bar{s}} \expval{g_s^2G^2} s^2}{294912 \pi ^6},\\
            \rho^{\langle\mathcal{O}_{11}\rangle}&=&-\frac{g_s ^2 m_s  \expval{q\bar{q}}^2 s (12  \expval{qG\bar{q}}+5 \expval{sG\bar{s}})}{663552 \pi ^6},\\
           \rho^{\langle\mathcal{O}_{12}\rangle}&=&-\frac{\expval{q\bar{q}}^4 s }{384 \pi ^2}+\frac{5 g_s ^4 \expval{q\bar{q}}^2 s \left(2 \expval{q\bar{q}}^2+\expval{s\bar{s}}^2\right)}{4478976 \pi ^6}-\frac{g_s ^2 \expval{q\bar{q}}^3 \expval{s\bar{s}} s}{10368 \pi ^4}.
     \end{eqnarray}

\subsection{$1^{+}$ $[3_c]_{qqq}$-$[\bar{3}_c]_{qqs}$}
\vspace{-1.5cm}
\begin{eqnarray}
         \rho^{\text{pert}}&=&\frac{s^7}{138726604800 \pi ^{10}},\\
          \rho^{\langle\mathcal{O}_3\rangle}&=&\frac{m_s s^5 (7 \expval{q\bar{q}}+3 \expval{s\bar{s}})}{275251200 \pi ^8},\\
          \rho^{\langle\mathcal{O}_4\rangle}&=&0,\\
           \rho^{\langle\mathcal{O}_5\rangle}&=&-\frac{m_s s^4 (18 \expval{qG\bar{q}}+5 \expval{sG\bar{s}})}{94371840 \pi ^8},\\
           \rho^{\langle\mathcal{O}_6\rangle}&=&\frac{\expval{q\bar{q}} s^4 (5 \expval{q\bar{q}}+3 \expval{s\bar{s}})}{2949120 \pi ^6}+\frac{g_s ^2 s^4 \left(5 \expval{q\bar{q}}^2+\expval{s\bar{s}}^2\right)}{127401984 \pi ^8},\\
           \rho^{\langle\mathcal{O}_7\rangle}&=&\frac{m_s  \expval{q\bar{q}} \expval{g_s^2G^2} s^3}{9437184 \pi ^8} ,\\
           \rho^{\langle\mathcal{O}_8\rangle}&=&\frac{s^3 (16 \expval{q\bar{q}} \expval{qG\bar{q}}+5 \expval{q\bar{q}} \expval{sG\bar{s}}+5 \expval{s\bar{s}} \expval{qG\bar{q}})}{983040 \pi ^6},\\
           \rho^{\langle\mathcal{O}_9\rangle}&=&-\frac{m_s \expval{q\bar{q}}^2 s^2 (8 \expval{q\bar{q}}+3 \expval{s\bar{s}})}{12288 \pi ^4}+\frac{g_s ^2 m_s  \expval{q\bar{q}}^2 s^2 (32 \expval{q\bar{q}}+15 \expval{s\bar{s}})}{2654208 \pi ^6},\\
            \rho^{\langle\mathcal{O}_{10}\rangle}&=&-\frac{\expval{q\bar{q}} \expval{g_s^2G^2} s^2 (3 \expval{q\bar{q}}+2 \expval{s\bar{s}})}{589824 \pi ^6}-\frac{\expval{qG\bar{q}} s^2 (3 \expval{qG\bar{q}}+2 \expval{sG\bar{s}})}{131072 \pi ^6},\\
            \rho^{\langle\mathcal{O}_{11}\rangle}&=&\frac{m_s \expval{q\bar{q}} s (27 \expval{q\bar{q}} \expval{qG\bar{q}}+2 \expval{q\bar{q}} \expval{sG\bar{s}}+6 \expval{s\bar{s}} \expval{qG\bar{q}})}{9216 \pi ^4}\\
            & &-\frac{g_s ^2 m_s  \expval{q\bar{q}}^2 s (18  \expval{qG\bar{q}}+5 \expval{sG\bar{s}})}{995328 \pi ^6},\\
           \rho^{\langle\mathcal{O}_{12}\rangle}&=&\frac{\expval{q\bar{q}}^3 s (\expval{q\bar{q}}+3 \expval{s\bar{s}})}{576 \pi ^2}+\frac{5 g_s ^4 \expval{q\bar{q}}^2 s \left(2 \expval{q\bar{q}}^2+\expval{s\bar{s}}^2\right)}{6718464 \pi ^6}\\
           & &-\frac{g_s ^2 \expval{q\bar{q}}^2 s \left(3 \expval{q\bar{q}}^2+3 \expval{q\bar{q}} \expval{s\bar{s}}+\expval{s\bar{s}}^2\right)}{31104 \pi ^4},\\
              \rho^{\langle\mathcal{O}_{13}\rangle}&=&-\frac{m_s \expval{qG\bar{q}} (4 \expval{q\bar{q}} (9 \expval{qG\bar{q}}+\expval{sG\bar{s}})+3 \expval{s\bar{s}} \expval{qG\bar{q}})}{24576 \pi ^4}.
     \end{eqnarray}

\subsection{$1^{+}$ $[3_c]_{qqq}$-$[\bar{3}_c]_{qsq}$}
\vspace{-1.5cm}
\begin{eqnarray}
         \rho^{\text{pert}}&=&\frac{s^7}{138726604800 \pi ^{10}},\\
          \rho^{\langle\mathcal{O}_3\rangle}&=&\frac{m_s s^5 (2 \expval{q\bar{q}}+\expval{s\bar{s}})}{91750400 \pi ^8},\\
          \rho^{\langle\mathcal{O}_4\rangle}&=&0,\\
           \rho^{\langle\mathcal{O}_5\rangle}&=&-\frac{m_s s^4 (3 \expval{qG\bar{q}}+ \expval{sG\bar{s}})}{18874368 \pi ^8},\\
           \rho^{\langle\mathcal{O}_6\rangle}&=&-\frac{\expval{q\bar{q}} s^4 (11 \expval{q\bar{q}}+5 \expval{s\bar{s}})}{5898240 \pi ^6}+\frac{g_s ^2 s^4 \left(5 \expval{q\bar{q}}^2+\expval{s\bar{s}}^2\right)}{127401984 \pi ^8},\\
           \rho^{\langle\mathcal{O}_7\rangle}&=&\frac{m_s  \expval{q\bar{q}} \expval{g_s^2G^2} s^3}{11796480 \pi ^8} ,\\
           \rho^{\langle\mathcal{O}_8\rangle}&=&\frac{s^3 (9 \expval{q\bar{q}} \expval{qG\bar{q}}+2 \expval{q\bar{q}} \expval{sG\bar{s}}+2 \expval{s\bar{s}} \expval{qG\bar{q}})}{491520 \pi ^6},\\
           \rho^{\langle\mathcal{O}_9\rangle}&=&-\frac{7m_s \expval{q\bar{q}}^2 s^2 (2 \expval{q\bar{q}}+ \expval{s\bar{s}})}{24576 \pi ^4}+\frac{g_s ^2 m_s  \expval{q\bar{q}}^2 s^2 (8 \expval{q\bar{q}}+5 \expval{s\bar{s}})}{884736 \pi ^6},\\
            \rho^{\langle\mathcal{O}_{10}\rangle}&=&-\frac{\expval{q\bar{q}} \expval{g_s^2G^2} s^2 (7 \expval{q\bar{q}}+3 \expval{s\bar{s}})}{1179648 \pi ^6}-\frac{\expval{qG\bar{q}} s^2 (7 \expval{qG\bar{q}}+3 \expval{sG\bar{s}})}{262144 \pi ^6},\\
            \rho^{\langle\mathcal{O}_{11}\rangle}&=&\frac{5 m_s \expval{q\bar{q}} s (9 \expval{q\bar{q}} \expval{qG\bar{q}}+ \expval{q\bar{q}} \expval{sG\bar{s}}+3 \expval{s\bar{s}} \expval{qG\bar{q}})}{18432 \pi ^4}\\
            & &-\frac{g_s ^2 m_s  \expval{q\bar{q}}^2 s (12  \expval{qG\bar{q}}+5 \expval{sG\bar{s}})}{995328 \pi ^6},\\
           \rho^{\langle\mathcal{O}_{12}\rangle}&=&\frac{\expval{q\bar{q}}^3 s (3 \expval{q\bar{q}}+5 \expval{s\bar{s}})}{1152 \pi ^2}+\frac{5 g_s ^4 \expval{q\bar{q}}^2 s \left(2 \expval{q\bar{q}}^2+\expval{s\bar{s}}^2\right)}{6718464 \pi ^6}\\
           & &-\frac{g_s ^2 \expval{q\bar{q}}^2 s \left(15 \expval{q\bar{q}}^2+8 \expval{q\bar{q}} \expval{s\bar{s}}+5 \expval{s\bar{s}}^2\right)}{124416 \pi ^4},\\
              \rho^{\langle\mathcal{O}_{13}\rangle}&=&-\frac{m_s \expval{qG\bar{q}} (2 \expval{q\bar{q}} (9 \expval{qG\bar{q}}+2\expval{sG\bar{s}})+3 \expval{s\bar{s}} \expval{qG\bar{q}})}{16384 \pi ^4}.
     \end{eqnarray}

\subsection{$1^{-}$ $[3_c]_{qqq}$-$[\bar{3}_c]_{qqs}$}
\vspace{-1.5cm}
\begin{eqnarray}
         \rho^{\text{pert}}&=&\frac{s^7}{138726604800 \pi ^{10}},\\
          \rho^{\langle\mathcal{O}_3\rangle}&=&-\frac{m_s s^5 (7 \expval{q\bar{q}}-3 \expval{s\bar{s}})}{275251200 \pi ^8},\\
          \rho^{\langle\mathcal{O}_4\rangle}&=&0,\\
           \rho^{\langle\mathcal{O}_5\rangle}&=&\frac{m_s s^4 (18 \expval{qG\bar{q}}-5 \expval{sG\bar{s}})}{94371840 \pi ^8},\\
           \rho^{\langle\mathcal{O}_6\rangle}&=&-\frac{\expval{q\bar{q}} s^4 (5 \expval{q\bar{q}}-3 \expval{s\bar{s}})}{2949120 \pi ^6}+\frac{g_s ^2 s^4 \left(5 \expval{q\bar{q}}^2+\expval{s\bar{s}}^2\right)}{127401984 \pi ^8},\\
           \rho^{\langle\mathcal{O}_7\rangle}&=&-\frac{m_s  \expval{q\bar{q}} \expval{g_s^2G^2} s^3}{9437184 \pi ^8},\\
           \rho^{\langle\mathcal{O}_8\rangle}&=&\frac{s^3 (16 \expval{q\bar{q}} \expval{qG\bar{q}}-5 \expval{q\bar{q}} \expval{sG\bar{s}}-5 \expval{s\bar{s}} \expval{qG\bar{q}})}{983040 \pi ^6},\\
           \rho^{\langle\mathcal{O}_9\rangle}&=&-\frac{m_s \expval{q\bar{q}}^2 s^2 (8 \expval{q\bar{q}}-3 \expval{s\bar{s}})}{12288 \pi ^4}+\frac{g_s ^2 m_s  \expval{q\bar{q}}^2 s^2 (15 \expval{s\bar{s}}-32 \expval{q\bar{q}})}{2654208 \pi ^6},\\
            \rho^{\langle\mathcal{O}_{10}\rangle}&=&-\frac{\expval{qG\bar{q}} s^2 (3 \expval{qG\bar{q}}-2 \expval{sG\bar{s}})}{131072 \pi ^6}-\frac{\expval{q\bar{q}} \expval{g_s^2G^2} s^2 (3 \expval{q\bar{q}}-2 \expval{s\bar{s}})}{589824 \pi ^6},\\
            \rho^{\langle\mathcal{O}_{11}\rangle}&=&-\frac{m_s \expval{q\bar{q}} s (27 \expval{q\bar{q}} \expval{qG\bar{q}}-2 \expval{q\bar{q}} \expval{sG\bar{s}}-6 \expval{s\bar{s}} \expval{qG\bar{q}})}{9216 \pi ^4}\\
            & &+\frac{g_s ^2 m_s  \expval{q\bar{q}}^2 s (18  \expval{qG\bar{q}}-5 \expval{sG\bar{s}})}{995328 \pi ^6},\\
           \rho^{\langle\mathcal{O}_{12}\rangle}&=&\frac{\expval{q\bar{q}}^3 s (\expval{q\bar{q}}-3 \expval{s\bar{s}})}{576 \pi ^2}-\frac{g_s ^2 \expval{q\bar{q}}^2 s \left(3 \expval{q\bar{q}}^2-3 \expval{q\bar{q}} \expval{s\bar{s}}+\expval{s\bar{s}}^2\right)}{31104 \pi ^4}\\
           & &+\frac{5 g_s ^4 \expval{q\bar{q}}^2 s \left(2 \expval{q\bar{q}}^2+\expval{s\bar{s}}^2\right)}{6718464 \pi ^6},\\
              \rho^{\langle\mathcal{O}_{13}\rangle}&=&\frac{m_s \expval{qG\bar{q}} (4 \expval{q\bar{q}} (9 \expval{qG\bar{q}}-\expval{sG\bar{s}})-3 \expval{s\bar{s}} \expval{qG\bar{q}})}{24576 \pi ^4}.
     \end{eqnarray}

\subsection{$1^{-}$ $[3_c]_{qqq}$-$[\bar{3}_c]_{qsq}$}
\vspace{-1.5cm}
\begin{eqnarray}
         \rho^{\text{pert}}&=&\frac{s^7}{138726604800 \pi ^{10}},\\
          \rho^{\langle\mathcal{O}_3\rangle}&=&\frac{m_s s^5 (2 \expval{q\bar{q}}+ \expval{s\bar{s}})}{91750400 \pi ^8},\\
          \rho^{\langle\mathcal{O}_4\rangle}&=&0,\\
           \rho^{\langle\mathcal{O}_5\rangle}&=&-\frac{m_s s^4 (3 \expval{qG\bar{q}}+ \expval{sG\bar{s}})}{18874368 \pi ^8},\\
           \rho^{\langle\mathcal{O}_6\rangle}&=&-\frac{\expval{q\bar{q}} s^4 ( \expval{q\bar{q}}-5 \expval{s\bar{s}})}{5898240 \pi ^6}+\frac{g_s ^2 s^4 \left(5 \expval{q\bar{q}}^2+\expval{s\bar{s}}^2\right)}{127401984 \pi ^8},\\
           \rho^{\langle\mathcal{O}_7\rangle}&=& \frac{m_s  \expval{q\bar{q}} \expval{g_s^2G^2} s^3}{11796480 \pi ^8},\\
           \rho^{\langle\mathcal{O}_8\rangle}&=&\frac{s^3 (-\expval{q\bar{q}} \expval{qG\bar{q}}+2 \expval{q\bar{q}} \expval{sG\bar{s}}+2 \expval{s\bar{s}} \expval{qG\bar{q}})}{491520 \pi ^6},\\
           \rho^{\langle\mathcal{O}_9\rangle}&=&\frac{ m_s \expval{q\bar{q}}^2 s^2 (2 \expval{q\bar{q}}+ \expval{s\bar{s}})}{24576 \pi ^4}+\frac{g_s ^2 m_s  \expval{q\bar{q}}^2 s^2 (8 \expval{q\bar{q}}+5 \expval{s\bar{s}})}{884736 \pi ^6},\\
            \rho^{\langle\mathcal{O}_{10}\rangle}&=&\frac{\expval{qG\bar{q}} s^2 (\expval{qG\bar{q}}-3 \expval{sG\bar{s}})}{262144 \pi ^6}+\frac{\expval{q\bar{q}} \expval{g_s^2G^2} s^2 (\expval{q\bar{q}}-3 \expval{s\bar{s}})}{1179648 \pi ^6},\\
            \rho^{\langle\mathcal{O}_{11}\rangle}&=&-\frac{m_s \expval{q\bar{q}} s (9 \expval{q\bar{q}} \expval{qG\bar{q}}+ \expval{q\bar{q}} \expval{sG\bar{s}}+3 \expval{s\bar{s}} \expval{qG\bar{q}})}{18432 \pi ^4}\\
            & &-\frac{g_s ^2 m_s  \expval{q\bar{q}}^2 s (12  \expval{qG\bar{q}}+5 \expval{sG\bar{s}})}{995328 \pi ^6},\\
           \rho^{\langle\mathcal{O}_{12}\rangle}&=&-\frac{\expval{q\bar{q}}^3 s (3 \expval{q\bar{q}}+ \expval{s\bar{s}})}{1152 \pi ^2}+\frac{5 g_s ^4 \expval{q\bar{q}}^2 s \left(2 \expval{q\bar{q}}^2+\expval{s\bar{s}}^2\right)}{6718464 \pi ^6}\\
           & &+\frac{g_s ^2 \expval{q\bar{q}}^2 s \left(3 \expval{q\bar{q}}^2-8 \expval{q\bar{q}} \expval{s\bar{s}}+\expval{s\bar{s}}^2\right)}{124416 \pi ^4},\\
              \rho^{\langle\mathcal{O}_{13}\rangle}&=&\frac{m_s \expval{qG\bar{q}} (2 \expval{q\bar{q}} (9 \expval{qG\bar{q}}+2\expval{sG\bar{s}})+3 \expval{s\bar{s}} \expval{qG\bar{q}})}{49152 \pi ^4}.
     \end{eqnarray}

\section{Spectra of Type-II Currents}

\subsection{$0^{-}$ $[3_c]_{qqq}$-$[\bar{3}_c]_{qqs}$}
\vspace{-1.5cm}
\begin{eqnarray}
         \rho^{\text{pert}}&=&\frac{s^7}{123312537600 \pi ^{10}},\\
          \rho^{\langle\mathcal{O}_3\rangle}&=&-\frac{m_s s^5 (2 \expval{q\bar{q}}-\expval{s\bar{s}})}{78643200 \pi ^8},\\
          \rho^{\langle\mathcal{O}_4\rangle}&=& 0,\\
           \rho^{\langle\mathcal{O}_5\rangle}&=&\frac{m_s s^4 (3 \expval{qG\bar{q}}-\expval{sG\bar{s}})}{15728640 \pi ^8},\\
           \rho^{\langle\mathcal{O}_6\rangle}&=&\frac{\expval{q\bar{q}} s^4 (2 \expval{q\bar{q}}+\expval{s\bar{s}})}{983040 \pi ^6}+\frac{g_s ^2 s^4 \left(5 \expval{q\bar{q}}^2+\expval{s\bar{s}}^2\right)}{106168320 \pi ^8},\\
           \rho^{\langle\mathcal{O}_7\rangle}&=&-\frac{m_s  \expval{q\bar{q}} \expval{g_s^2G^2} s^3}{9437184 \pi ^8},\\
           \rho^{\langle\mathcal{O}_8\rangle}&=&-\frac{s^3 (\expval{q\bar{q}} (4 \expval{qG\bar{q}}+\expval{sG\bar{s}})+\expval{s\bar{s}} \expval{qG\bar{q}})}{196608 \pi ^6},\\
           \rho^{\langle\mathcal{O}_9\rangle}&=&\frac{m_s \expval{q\bar{q}}^2 s^2 (2 \expval{q\bar{q}}-\expval{s\bar{s}})}{3072 \pi ^4}-\frac{g_s ^2 m_s  \expval{q\bar{q}}^2 s^2 (8 \expval{q\bar{q}}+5 \expval{s\bar{s}})}{663552 \pi ^6},\\
            \rho^{\langle\mathcal{O}_{10}\rangle}&=&\frac{\expval{qG\bar{q}} s^2 (2 \expval{qG\bar{q}}+\expval{sG\bar{s}})}{65536 \pi ^6}+\frac{\expval{q\bar{q}} \expval{g_s^2G^2} s^2 (2 \expval{q\bar{q}}-\expval{s\bar{s}})}{294912 \pi ^6},\\
            \rho^{\langle\mathcal{O}_{11}\rangle}&=&\frac{m_s \expval{q\bar{q}} s (\expval{q\bar{q}} (9 \expval{qG\bar{q}}-\expval{sG\bar{s}})-3 \expval{s\bar{s}} \expval{sG\bar{s}})}{3072 \pi ^4}\\
            & &+\frac{g_s ^2 m_s  \expval{q\bar{q}}^2 s (12  \expval{qG\bar{q}}+5 \expval{sG\bar{s}})}{663552 \pi ^6},\\
           \rho^{\langle\mathcal{O}_{12}\rangle}&=&\frac{\expval{q\bar{q}}^3 s (\expval{q\bar{q}}+2 \expval{s\bar{s}})}{384 \pi ^2}+\frac{5 g_s ^4 \expval{q\bar{q}}^2 s \left(2 \expval{q\bar{q}}^2+\expval{s\bar{s}}^2\right)}{4478976 \pi ^6}\\
           & &+\frac{g_s ^2 \expval{q\bar{q}}^2 s \left(3 \expval{q\bar{q}}^2+2 \expval{q\bar{q}} \expval{s\bar{s}}+\expval{s\bar{s}}^2\right)}{20736 \pi ^4},\\
              \rho^{\langle\mathcal{O}_{13}\rangle}&=&\frac{m_s \expval{qG\bar{q}} (-18 \expval{q\bar{q}} \expval{qG\bar{q}}+4 \expval{q\bar{q}} \expval{sG\bar{s}}+3 \expval{s\bar{s}} \expval{qG\bar{q}})}{12288 \pi ^4}.
     \end{eqnarray}

\subsection{$0^{+}$ $[3_c]_{qqq}$-$[\bar{3}_c]_{qqs}$}
\vspace{-1.5cm}
\begin{eqnarray}
         \rho^{\text{pert}}&=&\frac{s^7}{123312537600 \pi ^{10}},\\
          \rho^{\langle\mathcal{O}_3\rangle}&=&-\frac{m_s s^5 (2 \expval{q\bar{q}}+\expval{s\bar{s}})}{78643200 \pi ^8},\\
          \rho^{\langle\mathcal{O}_4\rangle}&=&0,\\
           \rho^{\langle\mathcal{O}_5\rangle}&=&-\frac{m_s s^4 (3 \expval{qG\bar{q}}+\expval{sG\bar{s}})}{15728640 \pi ^8},\\
           \rho^{\langle\mathcal{O}_6\rangle}&=&\frac{\expval{q\bar{q}} s^4 (2 \expval{q\bar{q}}-\expval{s\bar{s}})}{983040 \pi ^6}+\frac{g_s ^2 s^4 \left(5 \expval{q\bar{q}}^2+\expval{s\bar{s}}^2\right)}{106168320 \pi ^8},\\
           \rho^{\langle\mathcal{O}_7\rangle}&=&\frac{m_s  \expval{q\bar{q}} \expval{g_s^2G^2} s^3}{9437184 \pi ^8}\\
           \rho^{\langle\mathcal{O}_8\rangle}&=&\frac{s^3 (\expval{q\bar{q}} (-4 \expval{qG\bar{q}}+\expval{sG\bar{s}})+\expval{s\bar{s}} \expval{qG\bar{q}})}{196608 \pi ^6},\\
           \rho^{\langle\mathcal{O}_9\rangle}&=&\frac{m_s \expval{q\bar{q}}^2 s^2 (2 \expval{q\bar{q}}+\expval{s\bar{s}})}{3072 \pi ^4}+\frac{g_s ^2 m_s  \expval{q\bar{q}}^2 s^2 (8 \expval{q\bar{q}}+5 \expval{s\bar{s}})}{663552 \pi ^6},\\
            \rho^{\langle\mathcal{O}_{10}\rangle}&=&\frac{\expval{q\bar{q}} \expval{g_s^2G^2} s^2 (2 \expval{q\bar{q}}-\expval{s\bar{s}})}{294912 \pi ^6}+\frac{\expval{qG\bar{q}} s^2 (2 \expval{qG\bar{q}}-\expval{sG\bar{s}})}{65536 \pi ^6},\\
            \rho^{\langle\mathcal{O}_{11}\rangle}&=&-\frac{m_s \expval{q\bar{q}} s (\expval{q\bar{q}} (9 \expval{qG\bar{q}}+\expval{sG\bar{s}})+3 \expval{s\bar{s}} \expval{sG\bar{s}})}{3072 \pi ^4}\\
            & &-\frac{g_s ^2 m_s  \expval{q\bar{q}}^2 s (12  \expval{qG\bar{q}}+5 \expval{sG\bar{s}})}{663552 \pi ^6},\\
           \rho^{\langle\mathcal{O}_{12}\rangle}&=&\frac{\expval{q\bar{q}}^3 s (\expval{q\bar{q}}-2 \expval{s\bar{s}})}{384 \pi ^2}+\frac{5 g_s ^4 \expval{q\bar{q}}^2 s \left(2 \expval{q\bar{q}}^2+\expval{s\bar{s}}^2\right)}{4478976 \pi ^6}
           \\
           & &+\frac{g_s ^2 \expval{q\bar{q}}^2 s \left(3 \expval{q\bar{q}}^2-2 \expval{q\bar{q}} \expval{s\bar{s}}+\expval{s\bar{s}}^2\right)}{20736 \pi ^4},\\
              \rho^{\langle\mathcal{O}_{13}\rangle}&=&\frac{m_s \expval{qG\bar{q}} (18 \expval{q\bar{q}} \expval{qG\bar{q}}+4 \expval{q\bar{q}} \expval{sG\bar{s}}+3 \expval{s\bar{s}} \expval{qG\bar{q}})}{12288 \pi ^4}.
     \end{eqnarray}

\subsection{$0^{+}$ $[3_c]_{qqq}$-$[\bar{3}_c]_{qsq}$}
\vspace{-1.5cm}
\begin{eqnarray}
         \rho^{\text{pert}}&=&\frac{s^7}{123312537600 \pi ^{10}},\\
          \rho^{\langle\mathcal{O}_3\rangle}&=&-\frac{m_s s^5 (2 \expval{q\bar{q}}-\expval{s\bar{s}})}{78643200 \pi ^8},\\
          \rho^{\langle\mathcal{O}_4\rangle}&=& \rho^{\langle\mathcal{O}_{13}\rangle}=0,\\
           \rho^{\langle\mathcal{O}_5\rangle}&=&\frac{m_s s^4 (3 \expval{qG\bar{q}}-\expval{sG\bar{s}})}{15728640 \pi ^8},\\
           \rho^{\langle\mathcal{O}_6\rangle}&=&\frac{\expval{q\bar{q}}  \expval{s\bar{s}}s^4}{983040 \pi ^6}+\frac{g_s ^2 s^4 \left(5 \expval{q\bar{q}}^2+\expval{s\bar{s}}^2\right)}{106168320 \pi ^8},\\
           \rho^{\langle\mathcal{O}_7\rangle}&=&-\frac{m_s  \expval{q\bar{q}} \expval{g_s^2G^2} s^3}{9437184 \pi ^8},\\
           \rho^{\langle\mathcal{O}_8\rangle}&=&-\frac{s^3 \expval{q\bar{q}}\expval{sG\bar{s}}+\expval{s\bar{s}} \expval{qG\bar{q}})}{196608 \pi ^6},\\
           \rho^{\langle\mathcal{O}_9\rangle}&=&-\frac{g_s ^2 m_s  \expval{q\bar{q}}^2 s^2 (8 \expval{q\bar{q}}-5 \expval{s\bar{s}})}{663552 \pi ^6} ,\\
            \rho^{\langle\mathcal{O}_{10}\rangle}&=&\frac{s^2\expval{qG\bar{q}} \expval{sG\bar{s}}}{65536 \pi ^6},\\
            \rho^{\langle\mathcal{O}_{11}\rangle}&=& \frac{g_s ^2 m_s  \expval{q\bar{q}}^2 s (12  \expval{qG\bar{q}}-5 \expval{sG\bar{s}})}{663552 \pi ^6},\\
           \rho^{\langle\mathcal{O}_{12}\rangle}&=&-\frac{\expval{q\bar{q}}^4 s }{384 \pi ^2}+\frac{5 g_s ^4 \expval{q\bar{q}}^2 s \left(2 \expval{q\bar{q}}^2+\expval{s\bar{s}}^2\right)}{4478976 \pi ^6}+\frac{g_s ^2 \expval{q\bar{q}}^3 \expval{s\bar{s}} s}{10368 \pi ^4}.
     \end{eqnarray}

\subsection{$1^{-}$ $[3_c]_{qqq}$-$[\bar{3}_c]_{qqs}$}
\vspace{-1.5cm}
\begin{eqnarray}
         \rho^{\text{pert}}&=&\frac{s^7}{138726604800 \pi ^{10}},\\
          \rho^{\langle\mathcal{O}_3\rangle}&=&-\frac{m_s s^5 (7 \expval{q\bar{q}}-3 \expval{s\bar{s}})}{275251200 \pi ^8},\\
          \rho^{\langle\mathcal{O}_4\rangle}&=&0,\\
           \rho^{\langle\mathcal{O}_5\rangle}&=&\frac{m_s s^4 (18 \expval{qG\bar{q}}-5 \expval{sG\bar{s}})}{94371840 \pi ^8},\\
           \rho^{\langle\mathcal{O}_6\rangle}&=&\frac{\expval{q\bar{q}} s^4 (5 \expval{q\bar{q}}+3 \expval{s\bar{s}})}{2949120 \pi ^6}+\frac{g_s ^2 s^4 \left(5 \expval{q\bar{q}}^2+\expval{s\bar{s}}^2\right)}{127401984 \pi ^8},\\
           \rho^{\langle\mathcal{O}_7\rangle}&=&-\frac{m_s  \expval{q\bar{q}} \expval{g_s^2G^2} s^3}{9437184 \pi ^8} ,\\
           \rho^{\langle\mathcal{O}_8\rangle}&=&-\frac{s^3 (16 \expval{q\bar{q}} \expval{qG\bar{q}}+5 \expval{q\bar{q}} \expval{sG\bar{s}}+5 \expval{s\bar{s}} \expval{qG\bar{q}})}{983040 \pi ^6},\\
           \rho^{\langle\mathcal{O}_9\rangle}&=&-\frac{m_s \expval{q\bar{q}}^2 s^2 (8 \expval{q\bar{q}}-3 \expval{s\bar{s}})}{12288 \pi ^4}-\frac{g_s ^2 m_s  \expval{q\bar{q}}^2 s^2 (32 \expval{q\bar{q}}-15 \expval{s\bar{s}})}{2654208 \pi ^6},\\
            \rho^{\langle\mathcal{O}_{10}\rangle}&=&\frac{\expval{qG\bar{q}} s^2 (3 \expval{qG\bar{q}}+2 \expval{sG\bar{s}})}{131072 \pi ^6}+\frac{\expval{q\bar{q}} \expval{g_s^2G^2} s^2 (3 \expval{q\bar{q}}+2 \expval{s\bar{s}})}{589824 \pi ^6},\\
            \rho^{\langle\mathcal{O}_{11}\rangle}&=&\frac{m_s \expval{q\bar{q}} s (27 \expval{q\bar{q}} \expval{qG\bar{q}}-2 \expval{q\bar{q}} \expval{sG\bar{s}}-6 \expval{s\bar{s}} \expval{qG\bar{q}})}{9216 \pi ^4}\\
            & &+\frac{g_s ^2 m_s  \expval{q\bar{q}}^2 s (18  \expval{qG\bar{q}}-5 \expval{sG\bar{s}})}{995328 \pi ^6},\\
           \rho^{\langle\mathcal{O}_{12}\rangle}&=&\frac{\expval{q\bar{q}}^3 s (\expval{q\bar{q}}+3 \expval{s\bar{s}})}{576 \pi ^2}+\frac{5 g_s ^4 \expval{q\bar{q}}^2 s \left(2 \expval{q\bar{q}}^2+\expval{s\bar{s}}^2\right)}{6718464 \pi ^6}\\
           & &+\frac{g_s ^2 \expval{q\bar{q}}^2 s \left(3 \expval{q\bar{q}}^2+3 \expval{q\bar{q}} \expval{s\bar{s}}+\expval{s\bar{s}}^2\right)}{31104 \pi ^4},\\
              \rho^{\langle\mathcal{O}_{13}\rangle}&=&\frac{m_s \expval{qG\bar{q}} (4 \expval{q\bar{q}} (-9 \expval{qG\bar{q}}+\expval{sG\bar{s}})+3 \expval{s\bar{s}} \expval{qG\bar{q}})}{24576 \pi ^4}.
     \end{eqnarray}

\subsection{$1^{-}$ $[3_c]_{qqq}$-$[\bar{3}_c]_{qsq}$}
\vspace{-1.5cm}
\begin{eqnarray}
         \rho^{\text{pert}}&=&\frac{s^7}{138726604800 \pi ^{10}},\\
          \rho^{\langle\mathcal{O}_3\rangle}&=&-\frac{m_s s^5 (2 \expval{q\bar{q}}-\expval{s\bar{s}})}{91750400 \pi ^8},\\
          \rho^{\langle\mathcal{O}_4\rangle}&=&0,\\
           \rho^{\langle\mathcal{O}_5\rangle}&=&\frac{m_s s^4 (3 \expval{qG\bar{q}}- \expval{sG\bar{s}})}{18874368 \pi ^8},\\
           \rho^{\langle\mathcal{O}_6\rangle}&=&\frac{\expval{q\bar{q}} s^4 (11 \expval{q\bar{q}}+5 \expval{s\bar{s}})}{5898240 \pi ^6}+\frac{g_s ^2 s^4 \left(5 \expval{q\bar{q}}^2+\expval{s\bar{s}}^2\right)}{127401984 \pi ^8},\\
           \rho^{\langle\mathcal{O}_7\rangle}&=&-\frac{m_s  \expval{q\bar{q}} \expval{g_s^2G^2} s^3}{11796480 \pi ^8} ,\\
           \rho^{\langle\mathcal{O}_8\rangle}&=&-\frac{s^3 (9 \expval{q\bar{q}} \expval{qG\bar{q}}+2 \expval{q\bar{q}} \expval{sG\bar{s}}+2 \expval{s\bar{s}} \expval{qG\bar{q}})}{491520 \pi ^6},\\
           \rho^{\langle\mathcal{O}_9\rangle}&=&-\frac{7m_s \expval{q\bar{q}}^2 s^2 (2 \expval{q\bar{q}}- \expval{s\bar{s}})}{24576 \pi ^4}-\frac{g_s ^2 m_s  \expval{q\bar{q}}^2 s^2 (8 \expval{q\bar{q}}-5 \expval{s\bar{s}})}{884736 \pi ^6},\\
            \rho^{\langle\mathcal{O}_{10}\rangle}&=&\frac{\expval{qG\bar{q}} s^2 (7 \expval{qG\bar{q}}+3 \expval{sG\bar{s}})}{262144 \pi ^6}+\frac{\expval{q\bar{q}} \expval{g_s^2G^2} s^2 (7 \expval{q\bar{q}}+3 \expval{s\bar{s}})}{1179648 \pi ^6},\\
            \rho^{\langle\mathcal{O}_{11}\rangle}&=&\frac{5 m_s \expval{q\bar{q}} s (9 \expval{q\bar{q}} \expval{qG\bar{q}}- \expval{q\bar{q}} \expval{sG\bar{s}}-3 \expval{s\bar{s}} \expval{qG\bar{q}})}{18432 \pi ^4}\\
            & &+\frac{g_s ^2 m_s  \expval{q\bar{q}}^2 s (12  \expval{qG\bar{q}}-5 \expval{sG\bar{s}})}{995328 \pi ^6},\\
           \rho^{\langle\mathcal{O}_{12}\rangle}&=&\frac{\expval{q\bar{q}}^3 s (3 \expval{q\bar{q}}+5 \expval{s\bar{s}})}{1152 \pi ^2}+\frac{5 g_s ^4 \expval{q\bar{q}}^2 s \left(2 \expval{q\bar{q}}^2+\expval{s\bar{s}}^2\right)}{6718464 \pi ^6}\\
           & &+\frac{g_s ^2 \expval{q\bar{q}}^2 s \left(15 \expval{q\bar{q}}^2+8 \expval{q\bar{q}} \expval{s\bar{s}}+5 \expval{s\bar{s}}^2\right)}{124416 \pi ^4},\\
              \rho^{\langle\mathcal{O}_{13}\rangle}&=&\frac{m_s \expval{qG\bar{q}} (2 \expval{q\bar{q}} (-9 \expval{qG\bar{q}}+2\expval{sG\bar{s}})+3 \expval{s\bar{s}} \expval{qG\bar{q}})}{16384 \pi ^4}.
     \end{eqnarray}

\subsection{$1^{+}$ $[3_c]_{qqq}$-$[\bar{3}_c]_{qqs}$}
\vspace{-1.5cm}
\begin{eqnarray}
         \rho^{\text{pert}}&=&\frac{s^7}{138726604800 \pi ^{10}},\\
          \rho^{\langle\mathcal{O}_3\rangle}&=&\frac{m_s s^5 (7 \expval{q\bar{q}}+3 \expval{s\bar{s}})}{275251200 \pi ^8},\\
          \rho^{\langle\mathcal{O}_4\rangle}&=&0,\\
           \rho^{\langle\mathcal{O}_5\rangle}&=&-\frac{m_s s^4 (18 \expval{qG\bar{q}}+5 \expval{sG\bar{s}})}{94371840 \pi ^8},\\
           \rho^{\langle\mathcal{O}_6\rangle}&=&\frac{\expval{q\bar{q}} s^4 (5 \expval{q\bar{q}}-3 \expval{s\bar{s}})}{2949120 \pi ^6}+\frac{g_s ^2 s^4 \left(5 \expval{q\bar{q}}^2+\expval{s\bar{s}}^2\right)}{127401984 \pi ^8},\\
           \rho^{\langle\mathcal{O}_7\rangle}&=& \frac{m_s  \expval{q\bar{q}} \expval{g_s^2G^2} s^3}{9437184 \pi ^8},\\
           \rho^{\langle\mathcal{O}_8\rangle}&=&\frac{s^3 (-16 \expval{q\bar{q}} \expval{qG\bar{q}}+5 \expval{q\bar{q}} \expval{sG\bar{s}}+5 \expval{s\bar{s}} \expval{qG\bar{q}})}{983040 \pi ^6},\\
           \rho^{\langle\mathcal{O}_9\rangle}&=&\frac{m_s \expval{q\bar{q}}^2 s^2 (8 \expval{q\bar{q}}+3 \expval{s\bar{s}})}{12288 \pi ^4}+\frac{g_s ^2 m_s  \expval{q\bar{q}}^2 s^2 (32 \expval{q\bar{q}}+15 \expval{s\bar{s}})}{2654208 \pi ^6},\\
            \rho^{\langle\mathcal{O}_{10}\rangle}&=&\frac{\expval{qG\bar{q}} s^2 (3 \expval{qG\bar{q}}-2 \expval{sG\bar{s}})}{131072 \pi ^6}+\frac{\expval{q\bar{q}} \expval{g_s^2G^2} s^2 (3 \expval{q\bar{q}}-2 \expval{s\bar{s}})}{589824 \pi ^6},\\
            \rho^{\langle\mathcal{O}_{11}\rangle}&=&-\frac{m_s \expval{q\bar{q}} s (27 \expval{q\bar{q}} \expval{qG\bar{q}}+2 \expval{q\bar{q}} \expval{sG\bar{s}}+6 \expval{s\bar{s}} \expval{qG\bar{q}})}{9216 \pi ^4}\\
            & &-\frac{g_s ^2 m_s  \expval{q\bar{q}}^2 s (18  \expval{qG\bar{q}}+5 \expval{sG\bar{s}})}{995328 \pi ^6},\\
           \rho^{\langle\mathcal{O}_{12}\rangle}&=&\frac{\expval{q\bar{q}}^3 s (\expval{q\bar{q}}-3 \expval{s\bar{s}})}{576 \pi ^2}+\frac{5 g_s ^4 \expval{q\bar{q}}^2 s \left(2 \expval{q\bar{q}}^2+\expval{s\bar{s}}^2\right)}{6718464 \pi ^6}\\
           & &+\frac{g_s ^2 \expval{q\bar{q}}^2 s \left(3 \expval{q\bar{q}}^2-3 \expval{q\bar{q}} \expval{s\bar{s}}+\expval{s\bar{s}}^2\right)}{31104 \pi ^4},\\
              \rho^{\langle\mathcal{O}_{13}\rangle}&=&\frac{m_s \expval{qG\bar{q}} (4 \expval{q\bar{q}} (9 \expval{qG\bar{q}}+\expval{sG\bar{s}})+3 \expval{s\bar{s}} \expval{qG\bar{q}})}{24576 \pi ^4}.
     \end{eqnarray}

\subsection{$1^{+}$ $[3_c]_{qqq}$-$[\bar{3}_c]_{qsq}$}
\vspace{-1.5cm}
\begin{eqnarray}
         \rho^{\text{pert}}&=&\frac{s^7}{138726604800 \pi ^{10}},\\
          \rho^{\langle\mathcal{O}_3\rangle}&=&-\frac{m_s s^5 (2 \expval{q\bar{q}}+ \expval{s\bar{s}})}{91750400 \pi ^8},\\
          \rho^{\langle\mathcal{O}_4\rangle}&=&0,\\
           \rho^{\langle\mathcal{O}_5\rangle}&=&\frac{m_s s^4 (3 \expval{qG\bar{q}}- \expval{sG\bar{s}})}{18874368 \pi ^8},\\
           \rho^{\langle\mathcal{O}_6\rangle}&=&-\frac{\expval{q\bar{q}} s^4 ( \expval{q\bar{q}}-5 \expval{s\bar{s}})}{5898240 \pi ^6}+\frac{g_s ^2 s^4 \left(5 \expval{q\bar{q}}^2+\expval{s\bar{s}}^2\right)}{127401984 \pi ^8},\\
           \rho^{\langle\mathcal{O}_7\rangle}&=& -\frac{m_s  \expval{q\bar{q}} \expval{g_s^2G^2} s^3}{11796480 \pi ^8},\\
           \rho^{\langle\mathcal{O}_8\rangle}&=&\frac{s^3 (\expval{q\bar{q}} \expval{qG\bar{q}}-2 \expval{q\bar{q}} \expval{sG\bar{s}}-2 \expval{s\bar{s}} \expval{qG\bar{q}})}{491520 \pi ^6},\\
           \rho^{\langle\mathcal{O}_9\rangle}&=&\frac{ m_s \expval{q\bar{q}}^2 s^2 (2 \expval{q\bar{q}}- \expval{s\bar{s}})}{24576 \pi ^4}+\frac{g_s ^2 m_s  \expval{q\bar{q}}^2 s^2 (5 \expval{s\bar{s}}-8 \expval{q\bar{q}})}{884736 \pi ^6},\\
            \rho^{\langle\mathcal{O}_{10}\rangle}&=&-\frac{\expval{qG\bar{q}} s^2 (\expval{qG\bar{q}}-3 \expval{sG\bar{s}})}{262144 \pi ^6}-\frac{\expval{q\bar{q}} \expval{g_s^2G^2} s^2 (\expval{q\bar{q}}-3 \expval{s\bar{s}})}{1179648 \pi ^6},\\
            \rho^{\langle\mathcal{O}_{11}\rangle}&=&\frac{m_s \expval{q\bar{q}} s (-9 \expval{q\bar{q}} \expval{qG\bar{q}}+ \expval{q\bar{q}} \expval{sG\bar{s}}+3 \expval{s\bar{s}} \expval{qG\bar{q}})}{18432 \pi ^4}\\
            & &+\frac{g_s ^2 m_s  \expval{q\bar{q}}^2 s (12  \expval{qG\bar{q}}-5 \expval{sG\bar{s}})}{995328 \pi ^6},\\
           \rho^{\langle\mathcal{O}_{12}\rangle}&=&-\frac{\expval{q\bar{q}}^3 s (3 \expval{q\bar{q}}+ \expval{s\bar{s}})}{1152 \pi ^2}+\frac{5 g_s ^4 \expval{q\bar{q}}^2 s \left(2 \expval{q\bar{q}}^2+\expval{s\bar{s}}^2\right)}{6718464 \pi ^6}\\
           & &-\frac{g_s ^2 \expval{q\bar{q}}^2 s \left(3 \expval{q\bar{q}}^2-8 \expval{q\bar{q}} \expval{s\bar{s}}+\expval{s\bar{s}}^2\right)}{124416 \pi ^4},\\
              \rho^{\langle\mathcal{O}_{13}\rangle}&=&\frac{m_s \expval{qG\bar{q}} (2 \expval{q\bar{q}} (9 \expval{qG\bar{q}}-2\expval{sG\bar{s}})-3 \expval{s\bar{s}} \expval{qG\bar{q}})}{49152 \pi ^4}.
     \end{eqnarray}

\end{appendix}

\end{document}